\documentclass[aps,a4paper,twocolumn]{revtex4-2}

\usepackage{amsmath}
\usepackage{graphicx}
\usepackage{subfigure}
\usepackage[usenames,dvipsnames]{color}
\definecolor{darkblue}{RGB}{0,0,196}
\usepackage[colorlinks=true,linkcolor=darkblue,citecolor=darkblue,urlcolor=darkblue]{hyperref}
\usepackage{setspace}
\usepackage{multirow}
\usepackage{hyperref}
\usepackage{xcolor}
\hypersetup{
  colorlinks   = true, 
  urlcolor     = blue, 
  linkcolor    = blue, 
  citecolor   = blue 
}
\DeclareUnicodeCharacter{2212}{-}
\usepackage{graphicx}
\usepackage{amsmath,bbm}
\usepackage{amssymb,bm}
\usepackage{yfonts}
\usepackage{comment}
\usepackage[normalem]{ulem}

\begin{document}

\title{Characterizing nuclear modification effects in high-energy O-O collisions at energies available
at the CERN Large Hadron Collider: A transport model perspective}

\author{Debadatta Behera$^1$}
\email[]{debadatta3337@gmail.com}
\author{Suman Deb$^2$}
\email[]{sumandeb0101@gmail.com}
\author{Captain R. Singh$^1$}
\email[]{captainriturajsingh@gmail.com}
\author{Raghunath Sahoo$^1$}
\email[Corresponding author:]{Raghunath.Sahoo@cern.ch}

\affiliation{$^1$Department of Physics, Indian Institute of Technology Indore, Simrol, Indore 453552, India}
\affiliation{$^2$Laboratoire de Physique des 2 infinis Irène Joliot-Curie, Université Paris-Saclay, CNRS-IN2P3, F-91405 Orsay, France}

\begin{abstract}
The present work focuses on Oxygen-Oxygen (O-O) collisions, which are planned at the CERN Large Hadron Collider. Oxygen, being a doubly magic number nucleus, has some very unique features. This study attempts to probe the exotic state of QCD matter in O-O collisions. Additionally, the role of different nuclear density profiles in governing the final state dynamics in ultra-relativistic nuclear collisions is also explored. Using a multi-phase transport model, we obtain the nuclear modification factor ($\textit R_{\textit {AA}}$) for all charged hadrons and identified particles for O-O collisions at $\sqrt{s_{\rm{NN}}}$ = 7 TeV. Furthermore, we investigate the behavior of $\textit R_{\textit {AA}}$  as a function of transverse momentum ($\textit{p}_{\rm{T}}$) for three centralities (most central, mid-central, and peripheral) considering both $\alpha$-cluster and Woods-Saxon nuclear density profiles. We also extend this work to study the rapidity dependence of $\textit R_{\textit {AA}}$  for all charged hadrons. To better understand our findings of O-O collisions, the results are confronted with the available data of $\textit R_{\textit {AA}}$ for Pb-Pb collisions. The present study sheds light on particle production mechanisms, emphasizing factors influencing particle yield from pre-collision to post-collision stages in the context of O-O collisions.\\

\end{abstract}

 \maketitle


\section{Introduction}
\label{section1}

Ultra-relativistic hadronic and nuclear collisions at the Large Hadron Collider (LHC) have elevated 
physics to an entirely new level. Heavy-ion collisions at the LHC have facilitated the exploration of 
hot and dense QCD matter known as quark-gluon plasma (QGP). Hints of the existence of QCD matter, 
beyond heavy-ion collisions, have also been observed in the high multiplicity of $\textit{p}$-Pb and $\textit{pp}$ 
collisions at LHC energies~\cite{ALICE:2017jyt,Khachatryan:2016txc}. In this context, the future run 
at the LHC is expected to include a short run of O-O collisions~\cite{Brewer:2021kiv}. This brief 
run involving oxygen nuclei could provide a valuable opportunity to investigate the observed effects 
in high-multiplicity $\textit{p}$-Pb collisions, with a system having a relatively larger geometrical transverse overlap area but with a similar small number of participating nucleons and a similar number of final state multiplicity~\cite{ALICE:2021wim}. 
It is expected that a larger overlap area combined with approximately similar final-state multiplicity would enhance path-length-dependent effects such as jet quenching. Recently, several theoretical investigations such as those in Refs.~\cite{Lim:2018huo, Rybczynski:2019adt, Huang:2019tgz, Sievert:2019zjr, Schenke:2020mbo, Zakharov:2021uza, Huss:2020whe}, 
explored various aspects of particle production dynamics related to O-O collisions. In addition, O-O collisions provide an excellent opportunity to study the underlying mechanisms responsible for 
transverse collective flow effects, particle production, and light-nuclei production in a 
multiplicity range that bridges $\textit{pp}$ and $\textit{p}$-Pb on the lower side, and Xe-Xe and Pb-Pb on the higher 
side~\cite{ALICE:2021wim}.\\

Apart from these, the $\mathrm{^{16}{O}}$ has a reasonably compact structure and is resilient against decay 
due to its double magic property~\cite{Ropke:2017qck}. Furthermore, it is hypothesized that the 
$\alpha-$clustered structure~\cite{Li:2020vrg,Behera:2021zhi} has an additional impact on the oxygen
nuclei. A comprehensive analysis of the $\alpha-$clustered Oxygen nucleus is presented in Ref.~\cite{Behera:2021zhi}. Therefore, it is also important to investigate whether the initial nuclear structure affects the final-state observables, such as net particle production yield, the collective behavior of produced particles, etc. Thus, it is crucial to comprehend these observables as well as QGP-like properties while studying systems produced in O-O collisions.\\ 

Several studies have focused on understanding the characteristics of the quark-gluon plasma (QGP) in collisions of varying sizes, energies, and hence final state multiplicities. It has been shown in Ref~\cite{Sievert:2019zjr} that Pb-Pb collisions have a 60\% larger radius in comparison to O-O collisions at the same multiplicity. However, central O-O collisions exhibit significantly smaller 
eccentricities and higher temperatures than Pb-Pb collisions with the same multiplicity~\cite{Brewer:2021kiv}. Another interesting finding is that as the system size decreases, there is a more direct relationship 
between the initial eccentricity and the flow harmonics. To better understand QGP-like effects, it is crucial to study observable behavior in different collision systems with the same 
multiplicity at the LHC. This approach would greatly benefit experimental 
efforts. One possible way to investigate this idea is to compare observables sensitive to 
medium properties across different system sizes. In a nutshell, the investigation of the ultra-relativistic O-O collisions at the LHC energy would play a crucial role in the field of high-energy 
physics with the new notion of final state multiplicity driving the system properties while bridging the gap across relevant colliding species and centralities.\\
 
 In Sec.~\ref{section1}, we emphasized why O-O collisions are crucial. Further, the paper is organized into four sections. In Sec.~\ref{section2}, we delve into 
the event generation methodology and briefly present the formalism employed. The 
obtained results are thoroughly examined and discussed in Section~\ref{section3}. Finally, we 
conclude the paper by summarizing the main findings in Sec.~\ref{section4} and providing a 
potential outlook for future research. 

\section{Event Generation and Analysis Methodology}
\label{section2}

In this section, we give a brief overview of the AMPT model and discuss the  Woods-Saxon and  $\alpha-$clustered nuclear structure of the oxygen nucleus. At the end of this section, we have defined the nuclear modification factor, $\textit R_{\textit {AA}}$.

\subsection{A multi-phase transport (AMPT) model}
\label{subsection1}

A multi-phase transport model (AMPT) is a simulation framework for studying high-energy heavy-ion collisions. It consists of four 
main stages: initialization, parton cascade, hadronization, and hadron rescatterings~\cite{AMPT2, Zhang:2019utb}. In the initialization stage, 
the initial conditions for the collisions are generated using the HIJING model~\cite{ampthijing}. This includes information about 
the spatial and momentum distributions of minijet partons and soft string excitations. The parton cascade stage is carried out 
using Zhang's Parton Cascade (ZPC) model~\cite{Zhang:1997ej}. This stage simulates the interactions between partons and controls the 
partonic cross-section based on the values of the strong coupling constant  ($\alpha_{s}$) and the Debye screening mass ($\mu$).
Once the partons freeze out, the AMPT model uses a coalescence mechanism to combine them into hadrons; this process is called hadronization. After hadronization, the dynamics of the subsequent hadronic rescatterings are described by a relativistic transport (ART) model. This stage accounts for the interactions and collisions among the produced hadrons. In this particular study, we use the string melting mode of the AMPT model (version 2.26t9b) with Lund fragmentation parameters are $\textit{a}$ = 0.3, $\textit{b}$ = 0.15 $\rm GeV^{-2}$ and parton cross-section $\sigma_{gg} = \rm 3~mb$ for both oxygen-oxygen (O-O)~\cite{Behera:2021zhi, Lim:2018huo} and lead-lead (Pb-Pb) collisions. Lund  fragmentation parameters for proton-proton ($\textit{pp}$) collisions use $\textit{a}$ = 0.5 and $\textit{b}$ = 0.9 $\rm GeV^{-2}$ with $\sigma_{gg} = \rm 3~mb$~\cite{Lin:2021mdn}. Experimental data for $\textit{pp}$ collisions are utilized to compare the transverse momentum spectra~\cite{ALICE:2018vuu, ALICE:2013txf, ALICE:2015ial, ALICE:2016dei}.\\

In the context of heavy-ion collisions, the Woods-Saxon distribution is commonly used to represent the density profile of a nucleus. The Woods-Saxon charge density, based on a 3pF (three-parameter Fermi) distribution, is defined as;

\begin{equation}
\rho (r) = \frac{\rho_{0}(1+w(\frac{r}{r_0})^2)}{1+{ \rm exp}(\frac{r-r_0}{a})}.
\label{eq3}
\end{equation}


Here, $r_{0}$ is described as an equilibrium or saturation radius up to which nuclear matter is evenly distributed, and corresponding density distribution is defined by $\rho_{0}$ (a constant nuclear density at the core of a nucleus). While $\textit{r}$ is the radial distance, $\textit{w}$ is the deformation parameter, and $\textit{a}$ is the skin depth of the nucleus. For the oxygen nucleus, $r_{0}$ = 2.608 fm, $\textit{a}$ = 0.513 fm, and $\textit{w}$ = -0.051 according to Ref.~\cite{Li:2020vrg, Ding:2023ibq, Wang:2021ghq, Rybczynski:2017nrx,Svetlichnyi:2023nim}.\\

Recent theoretical studies have indicated the presence of $\alpha$-clustered structure in the $\mathrm{^{16}{O}}$--$\mathrm{^{16}{O}}$ nucleus~\cite{Li:2020vrg, Ding:2023ibq, Wang:2021ghq, Rybczynski:2017nrx,Svetlichnyi:2023nim}. In our study, we incorporated the $\alpha$-clustered structure into the oxygen nucleus using the AMPT model. This implementation was done numerically, where four $^4$He nuclei were positioned at the vertices of a regular tetrahedron structure. Each $^4$He nucleus follows the Woods-Saxon density profile described by Eq.~\ref{eq3}, with the parameters $r_0$ = 0.964 fm, $\textit{a}$ = 0.322 fm, and $\textit{w}$ = 0.517. The root mean square (rms) radius of $^4$He is determined to be 1.676 fm. The $\alpha$-clustered nuclei are arranged at the vertices of the tetrahedron, which has a side length of 3.42 fm. The root-mean-square (rms) radius of the oxygen nucleus is calculated to be 2.6999 fm. More details about the  $\alpha$-clustered implementation in the AMPT model can be found in Refs.~\cite{Behera:2023nwj, Behera:2021zhi}.\\

\begin{figure}[ht]
\includegraphics[scale=0.45]{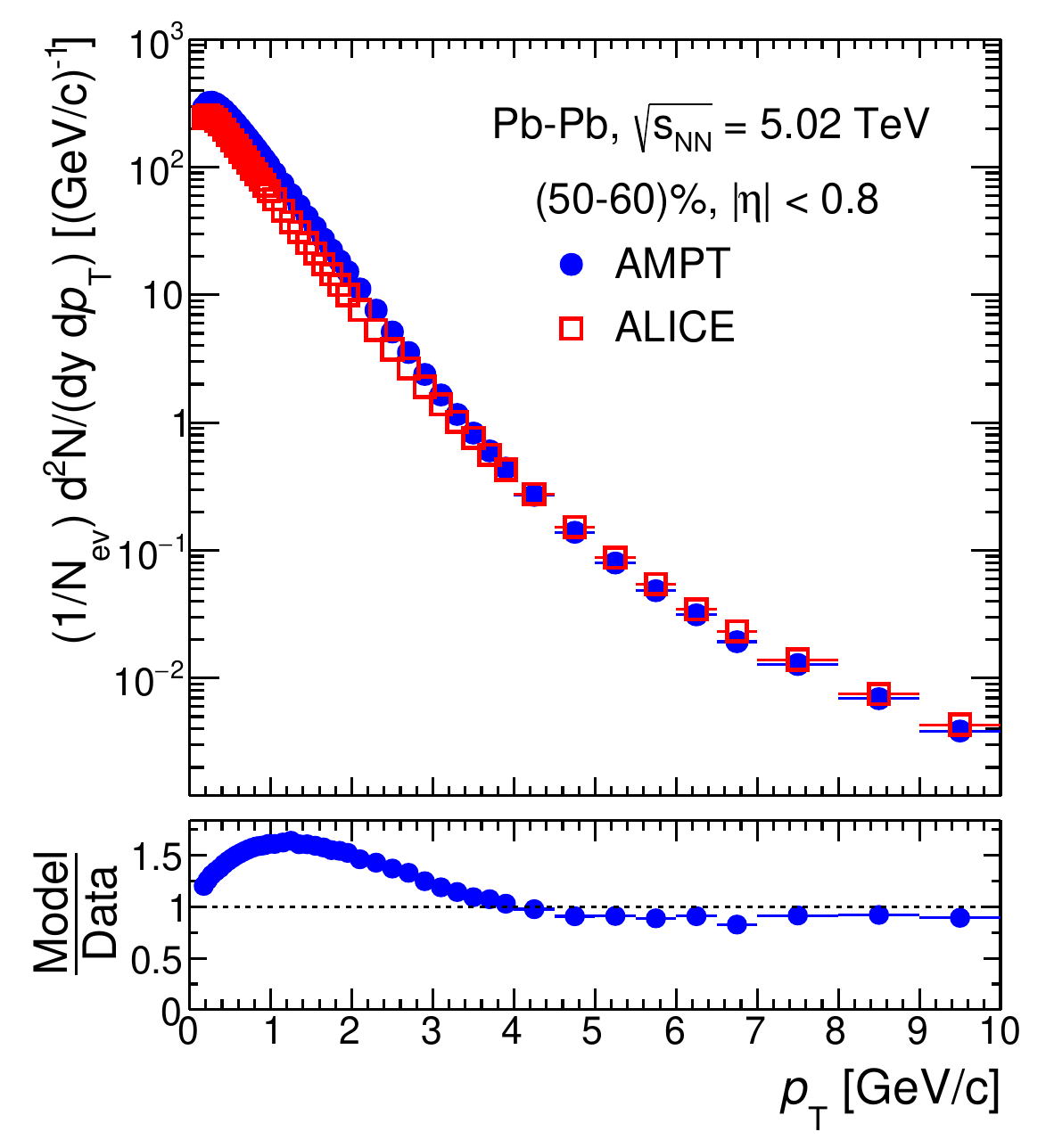}
\caption[]{(Color Online) Comparison of charged particles $\textit{p}_{\rm{T}}$ spectra obtained from AMPT for Pb-Pb collisions at $\sqrt{s_{\rm{NN}}}$ = 5.02 TeV with ALICE results~\cite{ALICE:2018vuu}. Statistical errors are within the markers.}
\label{pbpb50to60}
\end{figure}

\subsection{$\textit R_{\textit {AA}}$: Applicability and Comparison }
\label{subsection2}
In this work, we investigate the nuclear modification factor ($\textit R_{\textit {AA}}$) for all-charged hadrons and identified particles in O-O collisions at $\sqrt{s_{\rm{NN}}}$ = 
7 TeV. Table~\ref{table} presents number of binary collisions ($\langle N_{\rm{coll}} 
\rangle$) determined from Glauber model, number of participating nucleons ($\langle N_{\rm{part}} \rangle$) obtained from both Glauber and AMPT model and the average charged-particle multiplicity ($\langle{dN_{ch}}/d{\eta}\rangle$) from AMPT and ALICE~\cite{ALICE:2015juo} for O-O~\cite{Behera:2021zhi} and Pb-Pb collisions at $\sqrt{s_{\rm{NN}}}$ = 7 TeV and  5.02 TeV respectively, at mid-rapidity. The results displayed in table~\ref{table} indicate that the $\langle{dN_{ch}}/d{\eta}\rangle$ for the $\alpha$-clustered density 
profile for (0-5)\% of O-O collisions are approximately equivalent to the 
ALICE~\cite{ALICE:2015juo} Pb-Pb collisions for (50-60)\% centrality. This is a 
striking observation concerning the role of multiplicities in relation to 
thermalization in small systems, as studied by Landau in 1953~\cite{Landau:1965cpl}, 
and extended later in 1982 by van Hove, who explored the possibility of a quark-hadron phase transition in small systems using multiplicity as a 
probe~\cite{VanHove:1982vk}. This highlights the importance of multiplicities in the 
exploration of understanding the consequence of produced QCD-medium. Additionally, 
in our previous work~\cite{Behera:2021zhi}, we observed that the initial energy 
density produced in all collision centralities in O-O collisions are higher than the 
lattice QCD-predicted threshold for a deconfinement transition. This suggests the 
potential to create a QGP-like state even in oxygen nucleus collisions. Therefore, 
it is of interest to examine how the medium formed in two different colliding 
systems (O-O and Pb-Pb), which have different geometric overlap 
sizes~\cite{Sievert:2019zjr} but approximately similar numbers of final state 
charged particle multiplicity at specific centralities, affects particle production 
dynamics. However, an expected comparable $\langle{dN_{ch}}/d{\eta}\rangle$ for Pb-Pb collisions obtained from AMPT simulation and the estimation of ALICE experimental 
results show a discrepancy, as seen in Table~\ref{table}. This discrepancy could be 
attributed to the AMPT model's inadequacy to explain some experimental results, as 
discussed in refs.~\cite{ALICE:2013rdo,Lin:2021mdn}. Consequently, we proceed with 
studying the behaviors of $\textit R_{\textit {AA}}$ obtained from AMPT simulation of O-O 
collisions at (0-5)\% centrality and ALICE results of (50-60)\% centrality in Pb-Pb 
collisions, as multiplicities in these centralities are comparable. To provide a 
comprehensive and comparative conclusion, we consider both Woods-Saxon and $\alpha$-clustered nuclear density profiles for oxygen nuclei.
\\

\begin{table*}[htp]
   \begin{center}

                \caption{ Average charged particle multiplicity density ($\langle{dN_{ch}}/d{\eta}\rangle$) [ALICE~\cite{ALICE:2015juo} and AMPT], number of participating nucleons ($\langle N_{\rm{part}} \rangle$) (AMPT and Glauber) and number of binary collisons ($\langle N_{\rm{coll}} \rangle$) [Glauber] for O-O collisions~\cite{Behera:2021zhi} at $\sqrt{s_{\rm{NN}}}$ = 7 TeV and Pb-Pb system  $\sqrt{s_{\rm{NN}}}$ = 5.02 TeV in the range  $|\eta|<0.5$.}
                
                \label{tab:multiplicity}
                \hspace{1mm}
                 
\scalebox{0.95}
{
                \begin{tabular}{| c | c  |  c | c | c | c | c | c}

                \hline
               System & $\sqrt{s_{\rm{NN}}}$ (TeV) & Centrality (\%) &  $\langle N_{\rm{part}} \rangle ({\rm Glauber})$ &  $\langle N_{\rm{part}} \rangle$ (AMPT)& $\langle N_{\rm{coll}} \rangle$ (Glauber) & $\langle{dN_{ch}}/d{\eta}\rangle$  \\
                \hline

               \hline
                   O-O, $\alpha-$cluster & 7  & 0--5   & 29.43 $\pm$ 2.02 & 30.73 $\pm$ 2.06 & 55.12 $\pm$ 8.90         &187.54 $\pm$ 0.14     \\
                \hline
                    O-O, Woods-Saxon & 7  &  0--5    & 28.00 $\pm$ 2.06 & 29.26 $\pm$ 1.99 & 48.33 $\pm$  9.43     &161.07 $\pm$ 0.15     \\
                 \hline
 Pb-Pb & 5.02  & 50--60  & 53.6 $\pm$ 1.2 & 50.20 $\pm$ 7.04 & 90.88 $\pm$ 33.00 &183.00 $\pm$ 8.00 [ALICE]\\
 \cline{7-7} & & &&&& 235.25 $\pm$ 0.56 [AMPT] \\

                    \hline
                \end{tabular} 
                }   
            \label{table}
    \end{center}           
   
\end{table*}

The well known form of $\textit R_{\textit {AA}}$ is given as;\\

\begin{equation}
\label{eq2}
\rm{R_{AA}} = \frac {\rm d^{2}N^{{\rm AA}}/ d\rm{p_{T}}d\eta} {{\langle N_{coll} \rangle} {\rm d^{2}}N^{{\rm pp}}/d\rm{p_{T}}d\eta}
\end{equation}

where  $\textit N^{{\textit AA}}$ and $\textit N^{{\textit pp}}$ are the charged-particle yields in $\textit{A-A}$ and $\textit{pp}$ 
collisions respectively. The mean number of binary collisions is $\langle{\textit N_{coll}}\rangle  = \sigma^{\textit 
NN}_{\textit inel} \;\langle{\rm{T_{\rm {AA}}}}\rangle$, where $\sigma^{\textit NN}_{\textit inel}$ is the total 
inelastic nucleon-nucleon cross section and $\langle{\rm{T_{AA}}}\rangle$ is the mean nuclear thickness function. The $\langle{\textit N_{coll}}\rangle$ values for O-O collisions are taken from Ref.~\cite{Behera:2021zhi}, and for the Pb-Pb 
collisions, it is obtained via the impact parameter 
determination in the Glauber model~\cite{Miller:2007ri, 
Loizides:2016djv, Glauber:1970jm}.

\section{Results and Discussions}
\label{section3}

In this section, we first focus on evaluating the accuracy of the AMPT simulated 
data by comparing the transverse momentum ($\textit{p}_{\rm{T}}$) spectra in $\textit{pp}$ and Pb-Pb 
collisions with the corresponding ALICE experimental results. Next, we obtained the 
nuclear modification factor for charged hadrons in O-O collisions at 
$\sqrt{s_{\rm{NN}}}$ = 7 TeV. To examine the effect of scaling on the estimation of the nuclear modification factor in the work, we have used both ALICE experimental and AMPT simulated data for the respective yields from $\textit{pp}$ collisions. We define $\rm R^{Exp}_{AA}$ ($\rm R^{AMPT}_{AA}$) for the case where scaling is done using ALICE experimental (AMPT simulated) data. However, corresponding yields from O-O and Pb-Pb collisions are taken from the AMPT simulation only.
The influence of the Woods-Saxon and $\alpha$-clustered nuclear structures on the nuclear modification factor, taking into account identified particles, is also estimated. This observable, having dependence on fundamental quantities such as rapidity (\textit y), is very 
likely to be sensitive to the choice of the phase space. In order to take this fact into account, we have also investigated the dependence of the nuclear modification factor on the rapidity (\textit y) and centrality of the collisions.\\

\begin{figure}[ht]
\includegraphics[scale=0.45]{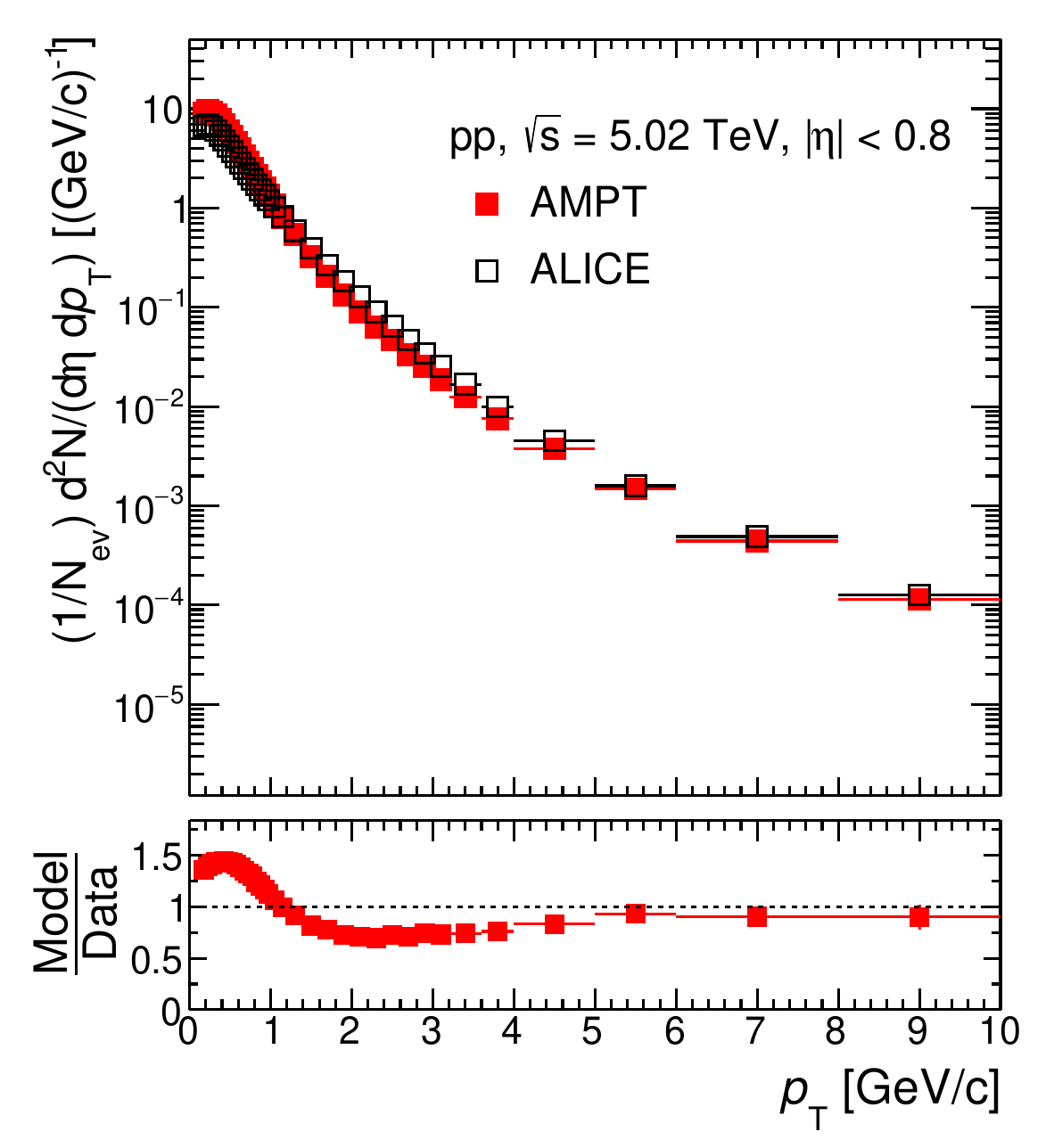}
\caption[]{(Color Online) Comparison of charged particles $\textit{p}_{\rm{T}}$ spectra 
obtained from AMPT for $\textit{pp}$ collisions at $\sqrt{s}$ = 5.02 TeV with ALICE 
results~\cite{ALICE:2018vuu}. Statistical errors are within the markers.}
\label{pp5tev}
\end{figure}

\begin{figure}[ht]
\includegraphics[scale=0.45]{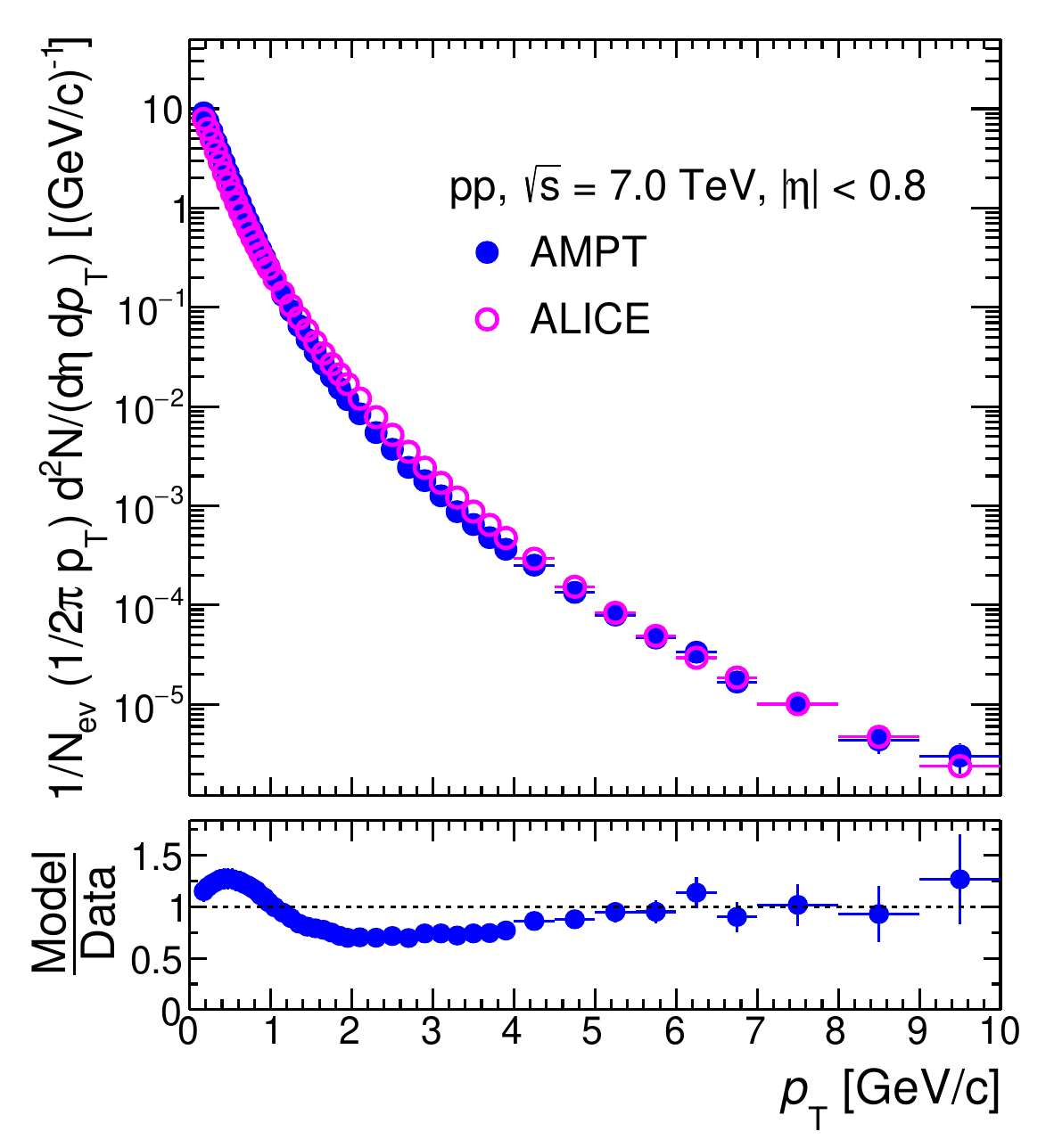}
\caption[]{(Color Online) Comparison of charged particles $\textit{p}_{\rm{T}}$ spectra 
obtained from AMPT for $\textit{pp}$ collisions at $\sqrt{s}$ = 7 TeV with ALICE 
results~\cite{ALICE:2013txf}. Statistical errors are within the markers.}
\label{pp7tev}
\end{figure}

\begin{figure}[ht]
\includegraphics[scale=0.45]{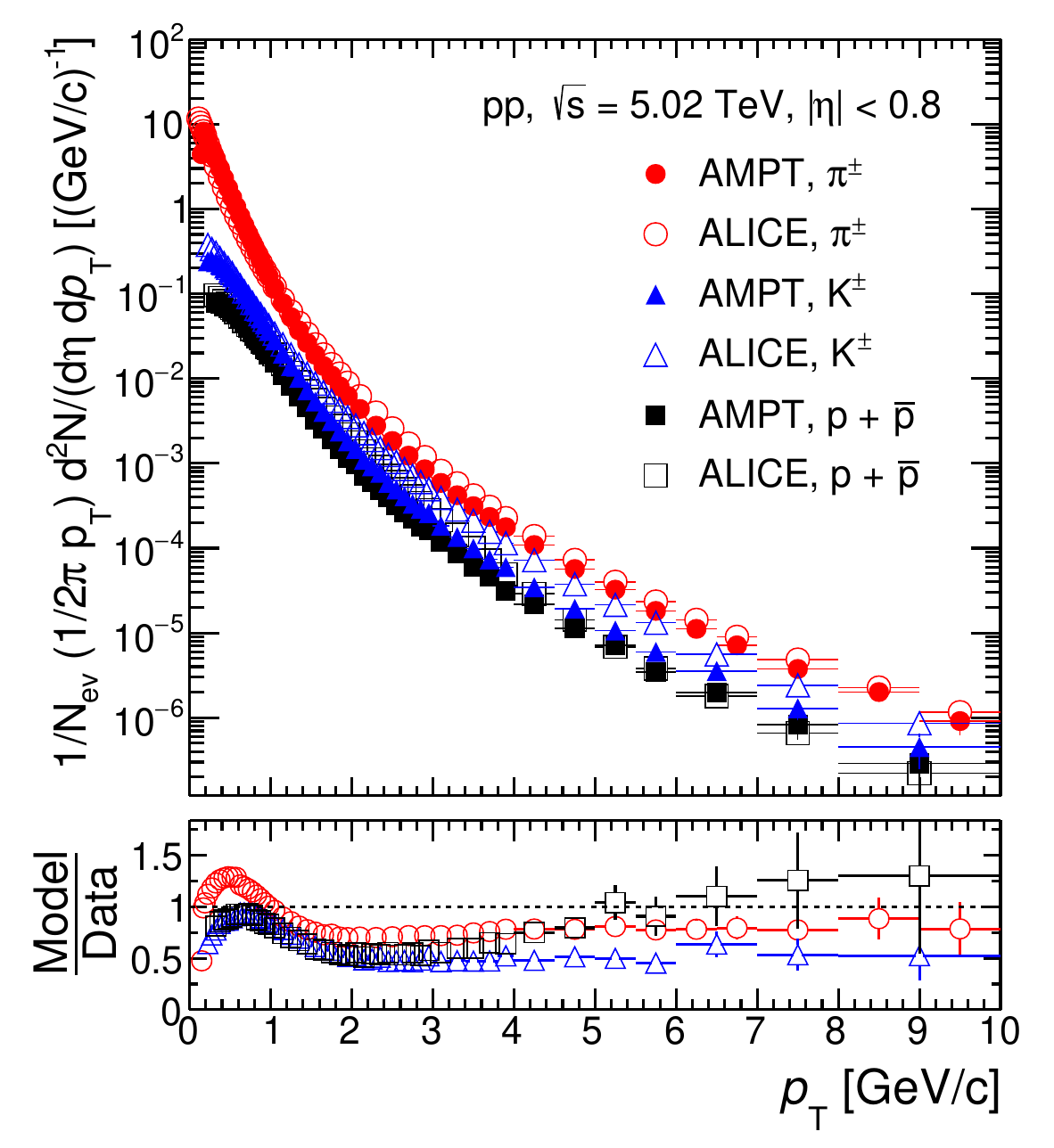}
\caption[]{(Color Online) Comparison of identified particles $\textit{p}_{\rm{T}}$ spectra 
obtained from AMPT for $\textit{pp}$ collisions at $\sqrt{s}$ = 5.02 TeV with ALICE 
results~\cite{ALICE:2016dei}. Statistical errors are within the markers.}
\label{pp5tevidentified}
\end{figure}

\begin{figure}[ht]
\includegraphics[scale=0.45]{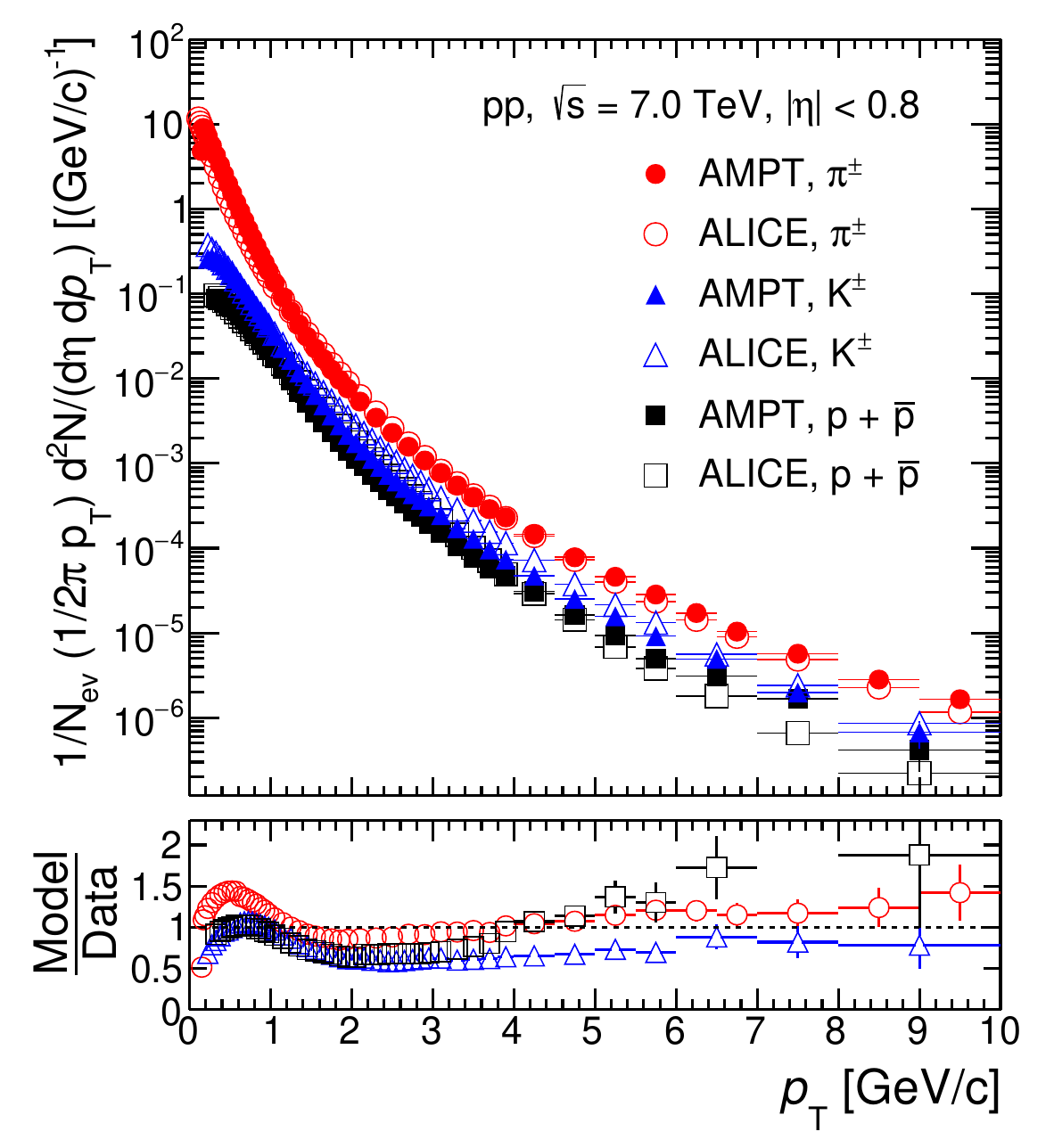}
\caption[]{(Color Online) Comparison of identified particles $\textit{p}_{\rm{T}}$ spectra 
obtained from AMPT for $\textit{pp}$ collisions at $\sqrt{s}$ = 7 TeV with ALICE 
results~\cite{ALICE:2015ial}. Statistical errors are within the markers.}
\label{pp7tevidentified}
\end{figure}

\begin{figure}[ht]
\includegraphics[scale=0.45]{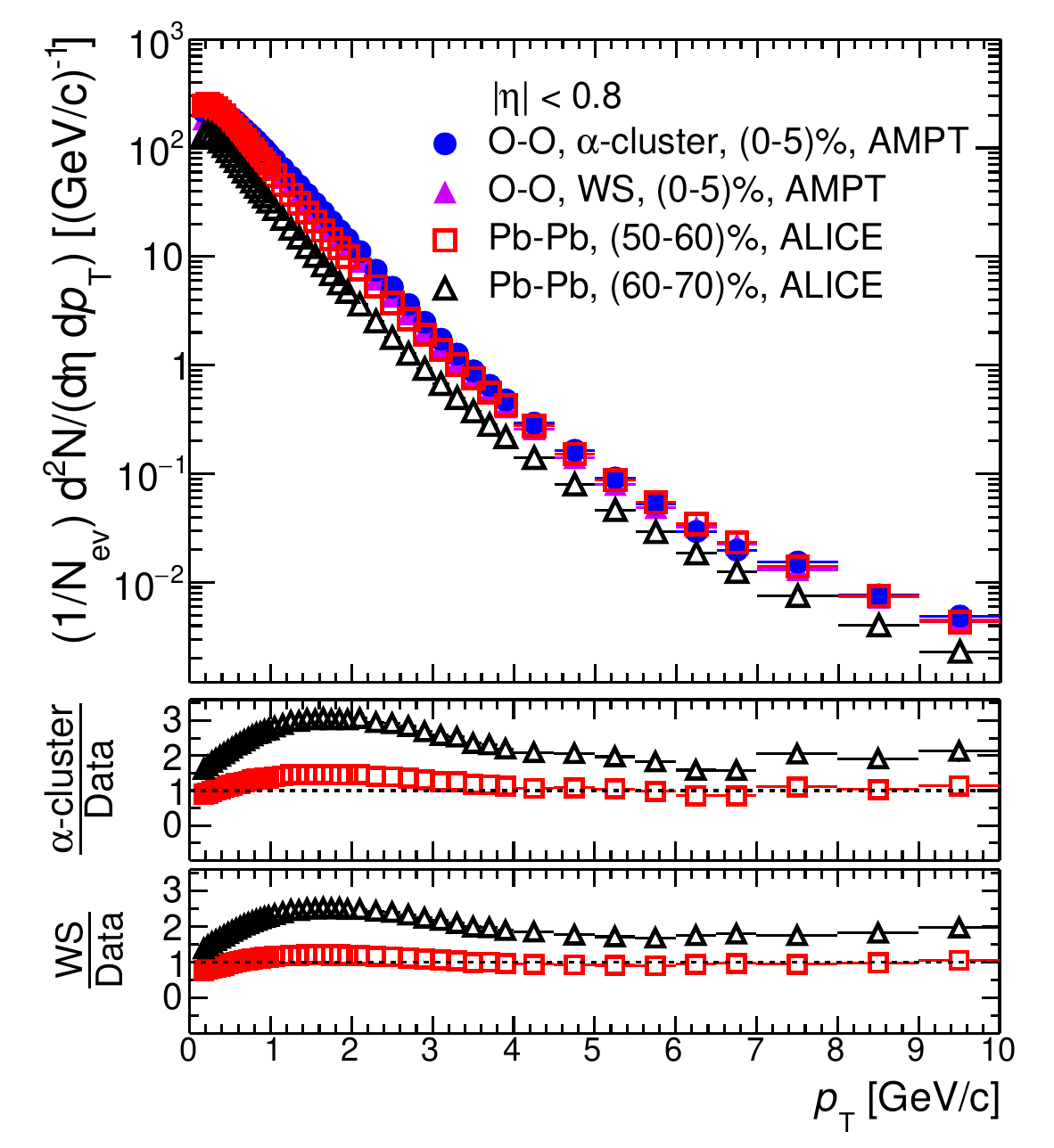}
\caption[]{(Color Online) Comparison of charged particles $\textit{p}_{\rm{T}}$ spectra of (0-
5)\% from AMPT result for O-O at $\sqrt{s_{\rm{NN}}}$ = 7 TeV with (50-60)\% and (60-
70)\% of ALICE results~\cite{ALICE:2018vuu} for Pb-Pb collisions at 
$\sqrt{s_{\rm{NN}}}$ = 5.02 TeV. Statistical errors are within the markers.}
\label{ptscomparision}
\end{figure}

\begin{figure}[ht]
\includegraphics[scale=0.45]{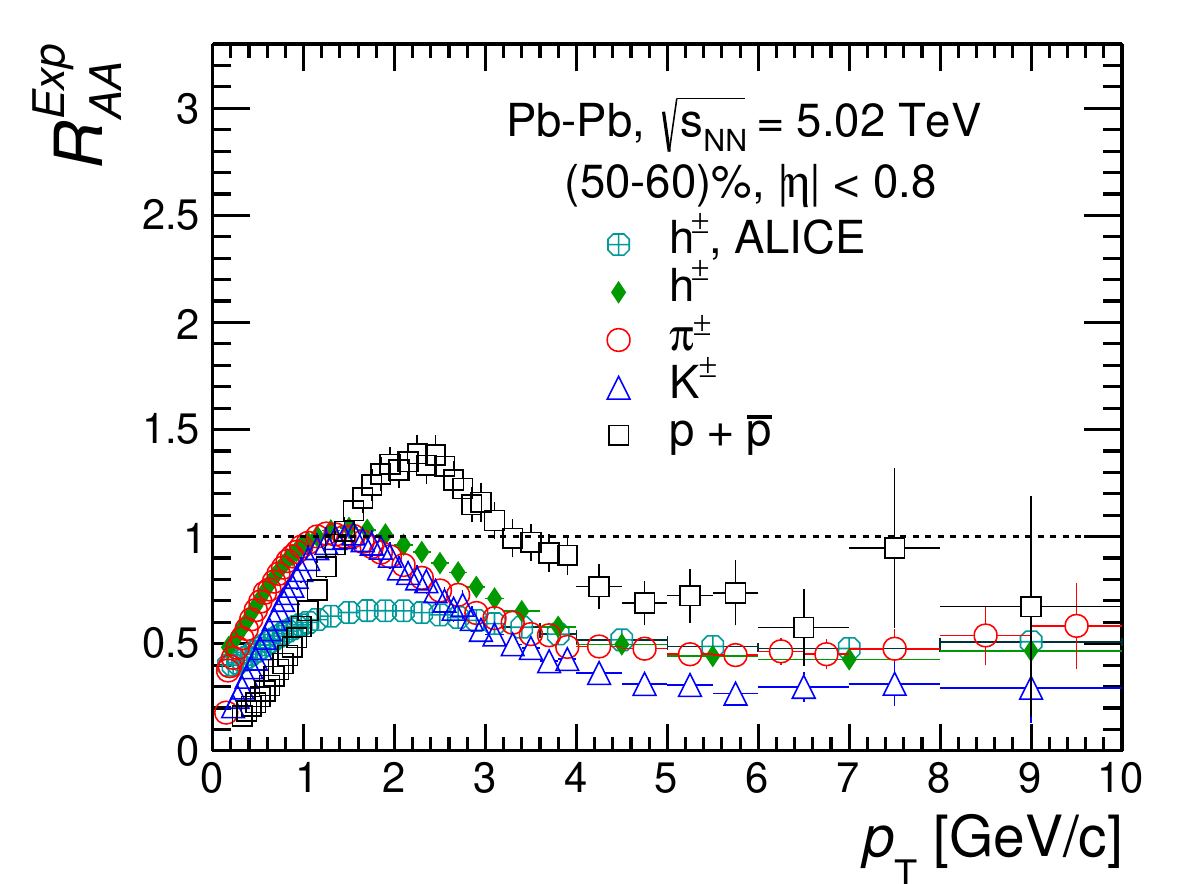}
\includegraphics[scale=0.45]{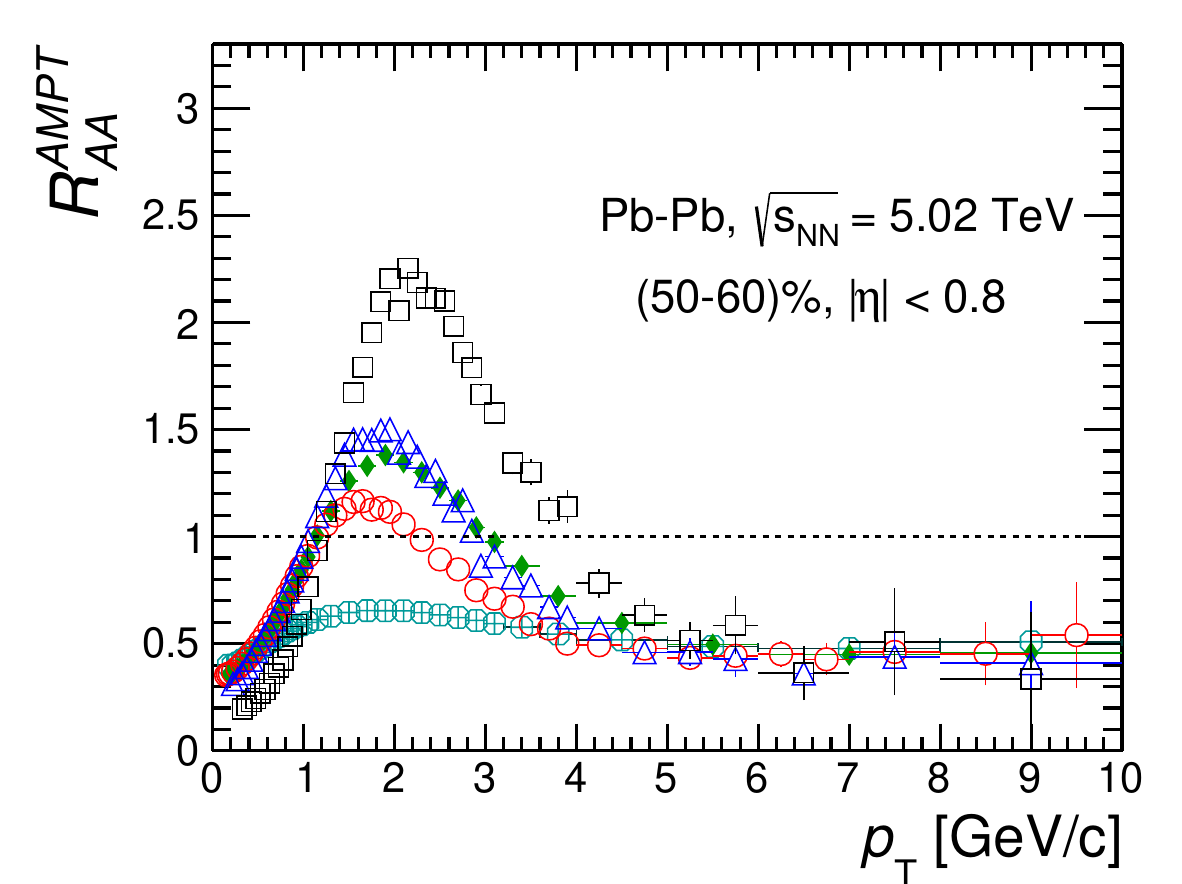}

\caption[]{(Color Online) Nuclear modification factor of charged hadrons and 
identified particles in Pb-Pb collisions at $\sqrt{s_{\rm{NN}}}$ = 5.02 TeV for (50-
60)\%, considering the $\textit{pp}$ yield from experimental data (ALICE) and  Pb-Pb yield from AMPT simulation (top). In the bottom panel, we show the nuclear modification considering both the $\textit{pp}$ and Pb-Pb yields from AMPT simulation ~\cite{ALICE:2018vuu}.}
\label{fig4}
\end{figure}

\subsection{Validation of the simulation via $\textit{p}_{\rm{T}}$ spectra}
The particle production yield and/or particle transverse momentum ($\textit{p}_{\rm{T}}$) 
spectra play a crucial role in exploring the particle production mechanisms in ultra-relativistic collisions. The study of high-$\textit{p}_{\rm{T}}$ particle production and 
parton energy loss provide insights into the dynamics of QCD matter. The partons 
experience energy loss as they traverse through a medium, which causes the parton 
splitting and gluon emission. Understanding the mechanism of parton energy loss is 
thus one of the primary goals of heavy-ion collisions. The energy loss can be 
determined by comparing the $\textit{p}_{\rm{T}}$ spectra of pp ($\textit{pp}$) collisions 
with heavy-ion collisions at the same energy. Understanding the precise method by 
which hard particles lose energy while traversing through the medium requires a 
thorough investigation of $\textit R_{\textit {AA}}$. However, estimation of this observable 
requires input from both $\textit{pp}$ and heavy-ion collisions as shown in Eq.~\ref{eq2} and 
thus, first of all, $\textit{p}_{\rm{T}}$ spectra of all charged particles obtained from AMPT 
are compared with ALICE results, to validate the simulation tuning. 
Figure~\ref{pbpb50to60} shows this comparison in Pb-Pb collisions at 
$\sqrt{s_{\rm{NN}}}$ = 5.02 TeV for (50-60)\% centrality, and Fig.~\ref{pp5tev} 
(\ref{pp7tev}) shows the comparison of charged particles $\textit{p}_{\rm{T}}$ spectra in $\textit{pp}$ 
collisions at $\sqrt{s}$ = 5.02 TeV (7 TeV) for minimum bias. This comparison 
reveals that the spectral shape from AMPT agrees well with the experimental data 
towards high-$\textit{p}_{\rm{T}}$ ($>$ 4 GeV), while there is a reasonable degree of 
deviation from the experimental results towards low-$\textit{p}_{\rm{T}}$ range. However, 
within uncertainties, this also validates that the tuning of the AMPT simulation 
used in this work quantitatively matches the experimental results within the 
acceptance range. Figure~\ref{pp5tevidentified} (\ref{pp7tevidentified}) shows  the 
comparison of $\textit{p}_{\rm{T}}$ spectra of identified particles obtained from both the 
ALICE experimental~\cite{ALICE:2016dei, ALICE:2015ial} and AMPT simulated data at 
$\sqrt{s}$ = 5.02 (7) TeV. These comparisons reveal that, when accounting for 
uncertainties, pions from AMPT exhibit relatively good agreement with the 
experimental data, particularly in the high-$\textit{p}_{\rm{T}}$ region. However, both kaons 
and protons exhibit significant deviations. As we shift towards low-$\textit{p}_{\rm{T}}$ 
ranges ($<1$ GeV), it becomes evident that kaons and protons align more closely with 
the experimental data, while a noticeable discrepancy remains in the case of pions. 
It is worth mentioning here that the scaling of $\textit R_{\textit {AA}}$ in the rest of the 
paper is done considering $\textit{p}_{\rm{T}}$ spectra in $\textit{pp}$ collisions from  both 
ALICE~\cite{ALICE:2015ial, ALICE:2018vuu, ALICE:2016dei, ALICE:2013txf} and AMPT.


\begin{figure*}[ht!]
\includegraphics[scale=0.295]{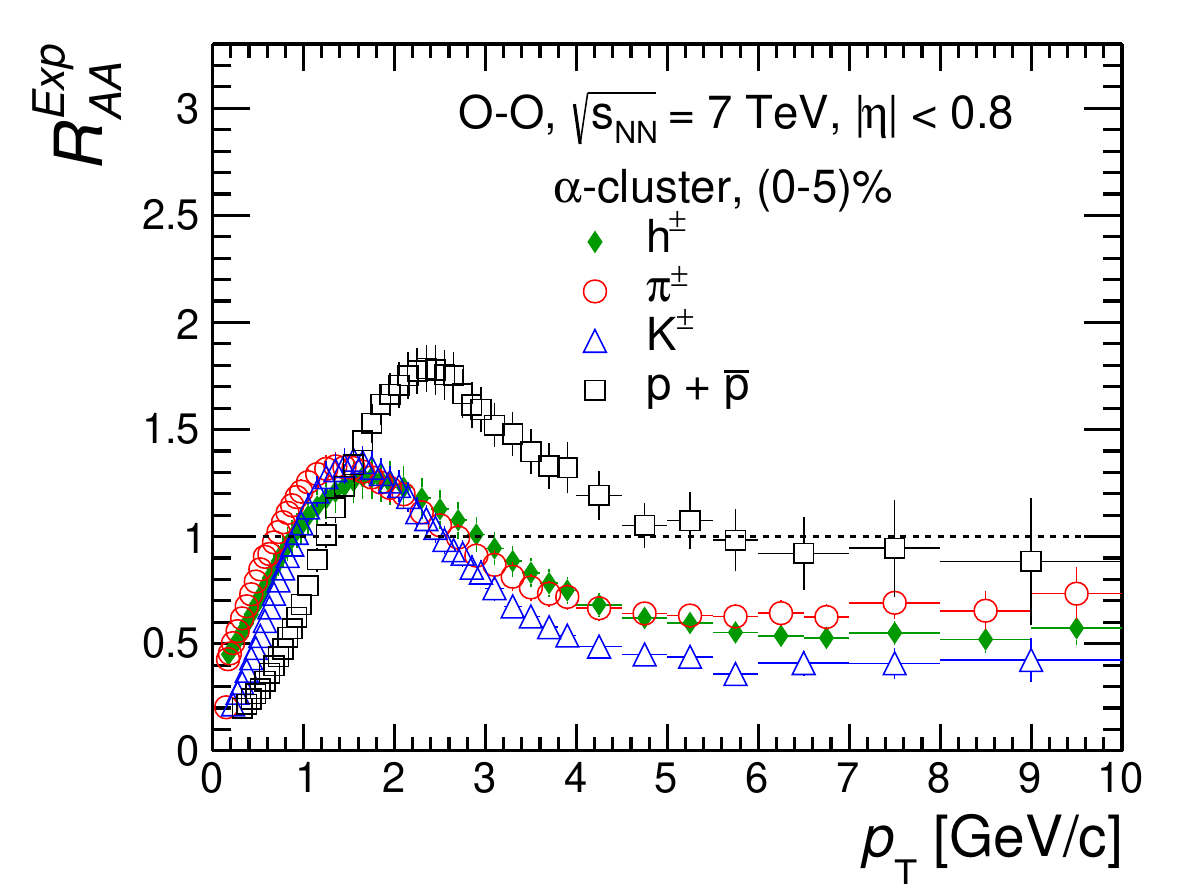}
\includegraphics[scale=0.295]{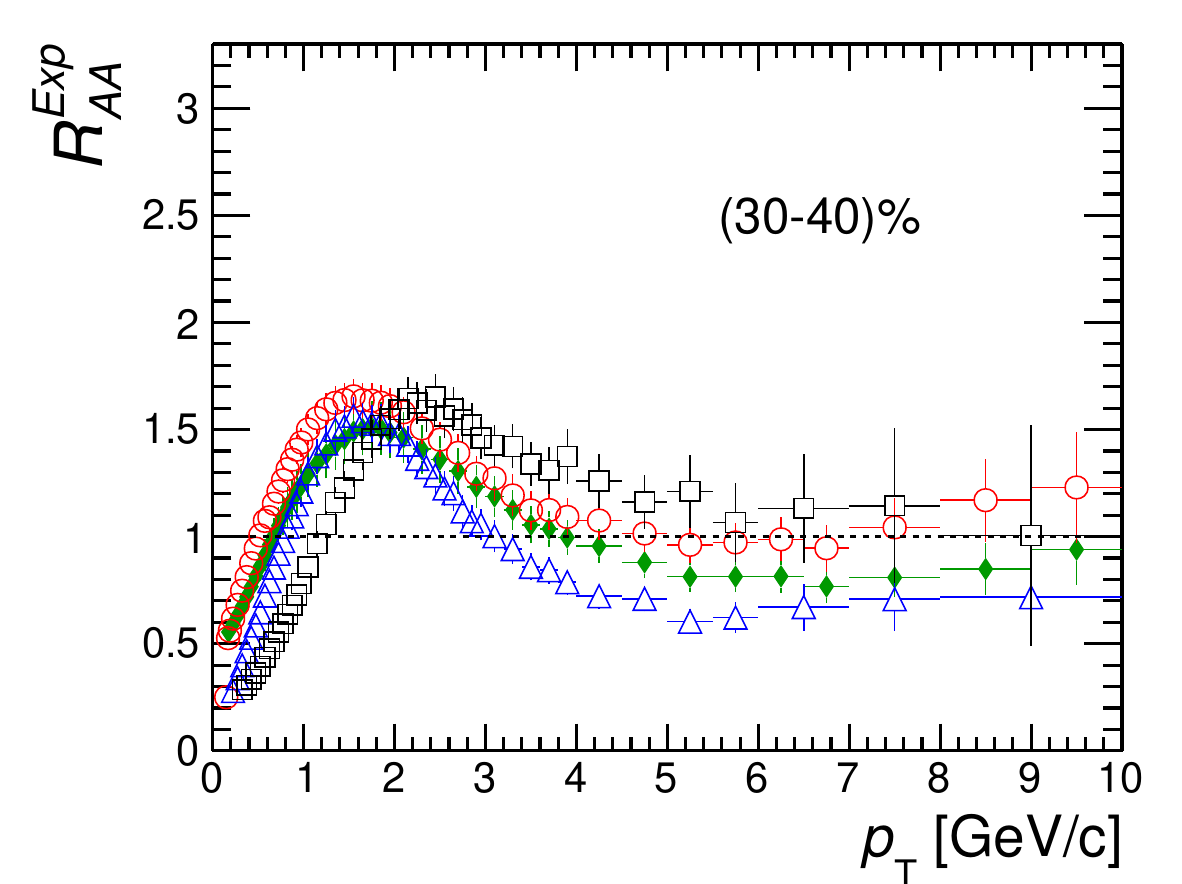}
\includegraphics[scale=0.295]{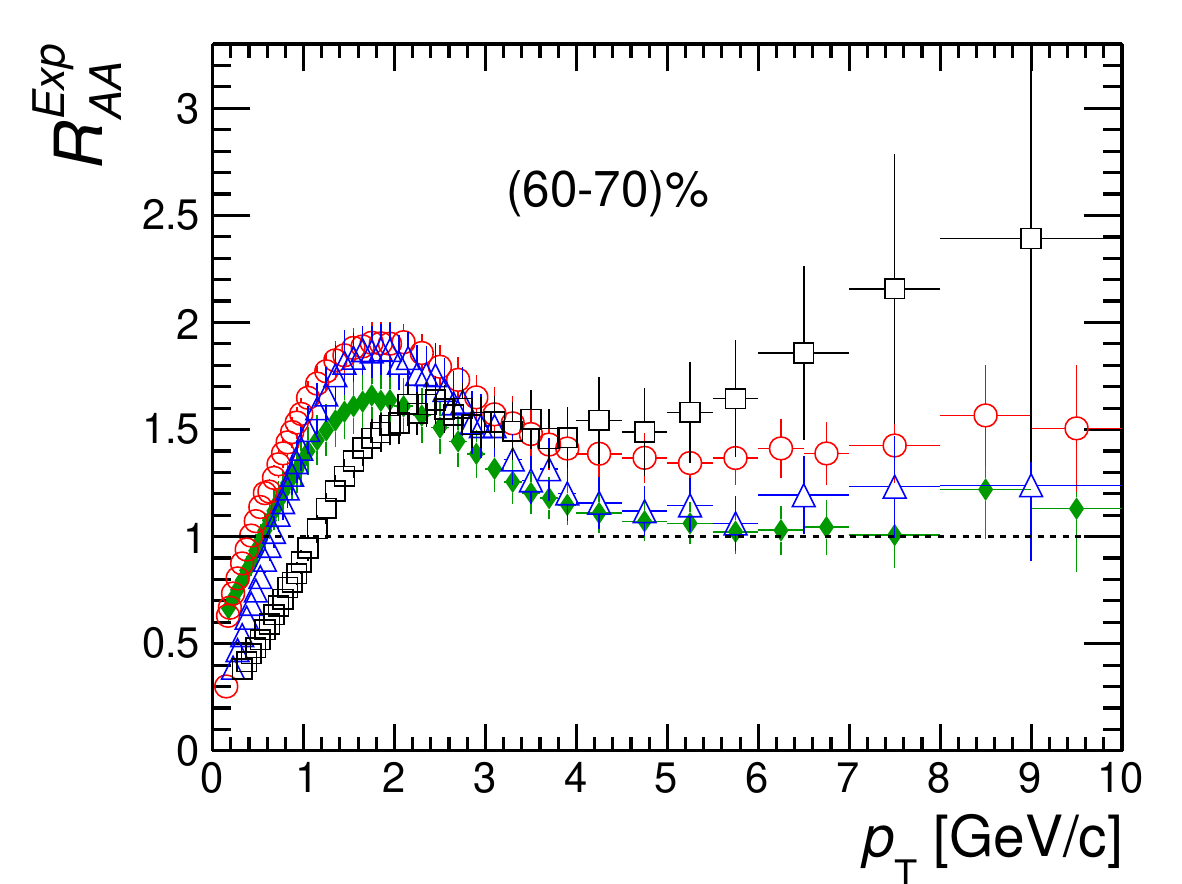}
\caption[]{(Color Online) {Nuclear modification factor ($\textit R_{\textit {AA}}$) of all 
charged hadrons ($\rm{h}^{\pm}$) and identified particles ($\pi^{\pm}$, 
$\rm{K}^{\pm}$ and $\rm{p}+\bar\rm{p}$) in O-O collisions at $\sqrt{s_{\rm{NN}}}$ = 
7 TeV for (0-5)\% [left], (30-40)\% [middle] and (60-70)\% [right], with $\alpha$-
clustered nuclear structure, taking $\textit{p}_{\rm{T}}$ spectra of $\textit{pp}$ from ALICE for 
estimating $\textit R_{\textit {AA}}$.} }
\label{fig2}
\end{figure*}

\begin{figure*}[ht!]
\includegraphics[scale=0.295]{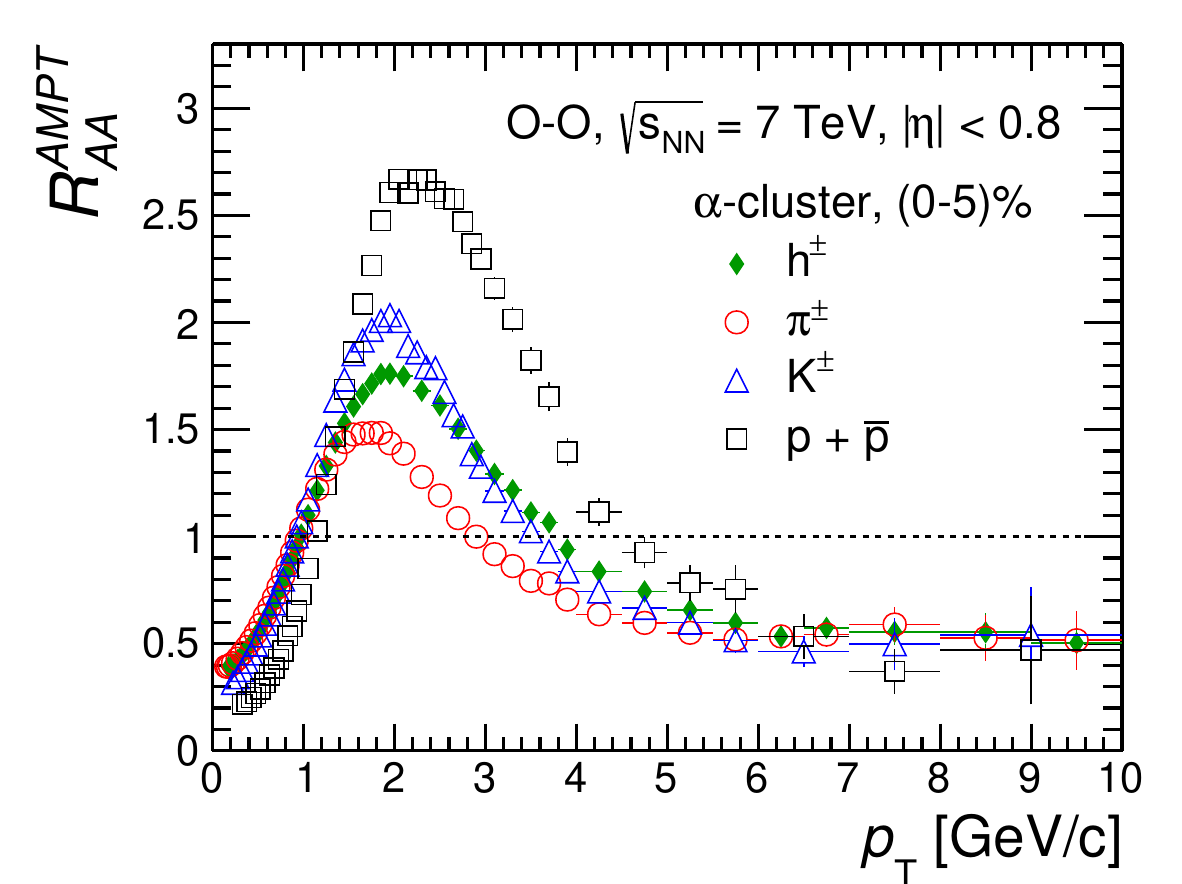}
\includegraphics[scale=0.295]{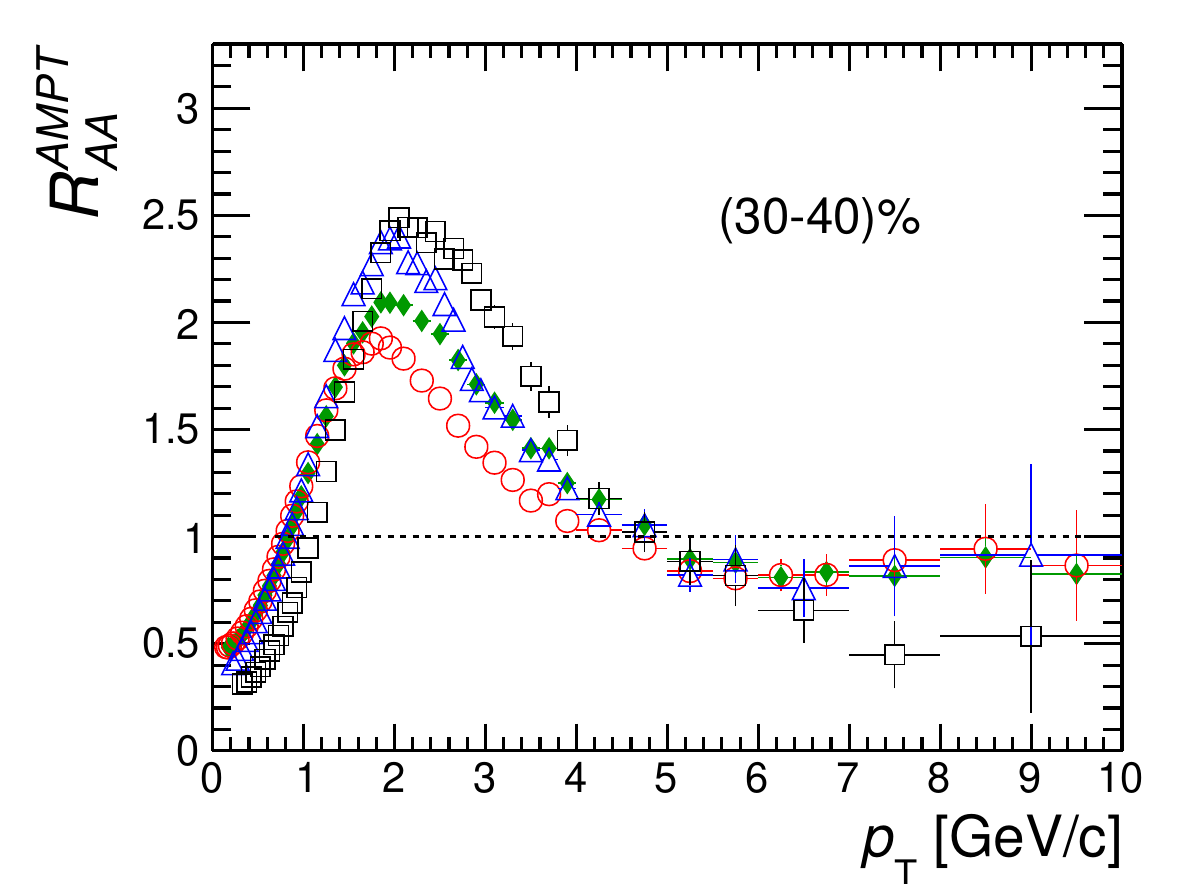}
\includegraphics[scale=0.295]{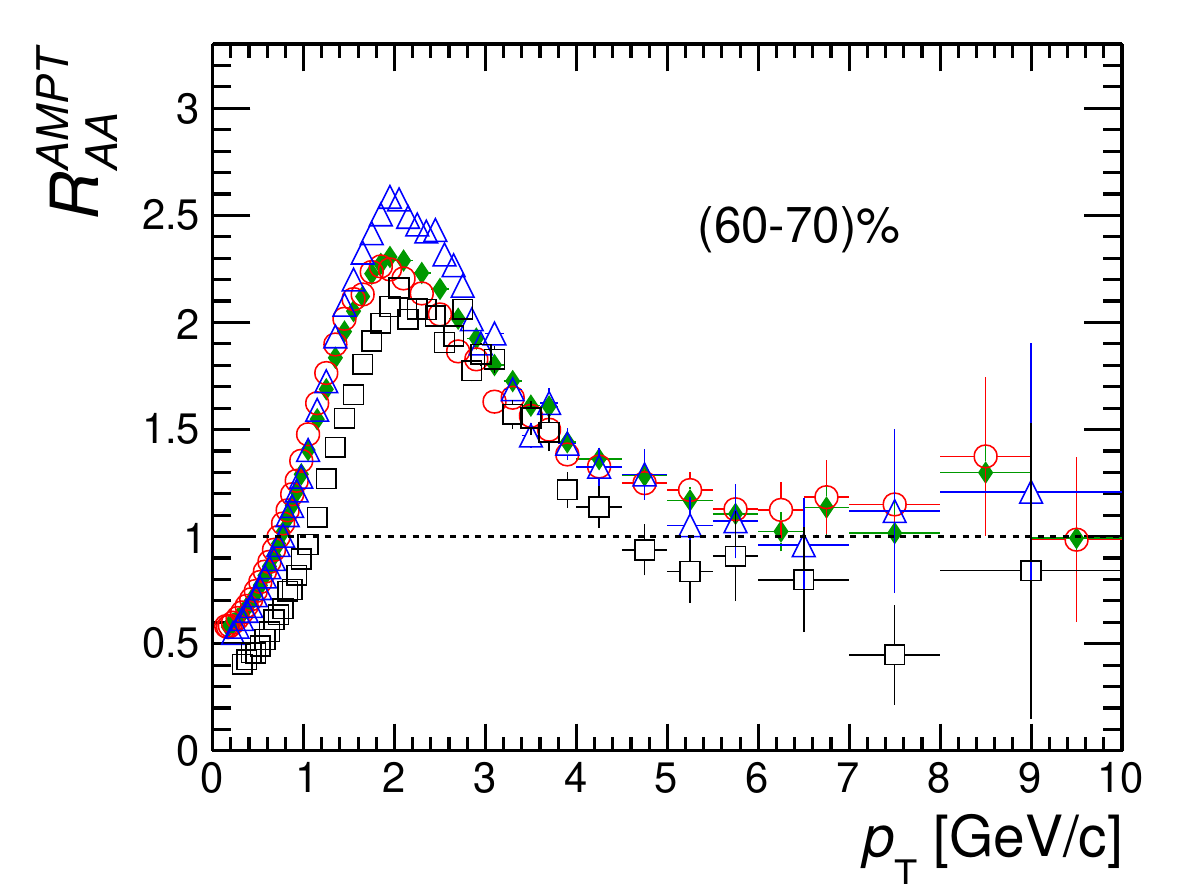}
\caption[]{(Color Online) {Nuclear modification factor ($\textit R_{\textit {AA}}$) of all 
charged hadrons ($\rm{h}^{\pm}$) and identified particles ($\pi^{\pm}$, 
$\rm{K}^{\pm}$ and $\rm{p}+\bar\rm{p}$) in O-O collisions at $\sqrt{s_{\rm{NN}}}$ = 
7 TeV for (0-5)\% [left], (30-40)\% [middle] and (60-70)\% [right], with $\alpha$-
clustered nuclear structure, taking $\textit{p}_{\rm{T}}$ spectra of $\textit{pp}$ from AMPT for estimating 
$\textit R_{\textit {AA}}$. } }
\label{fig2ampt}
\end{figure*}

\begin{figure*}[ht!]
\includegraphics[scale=0.295]{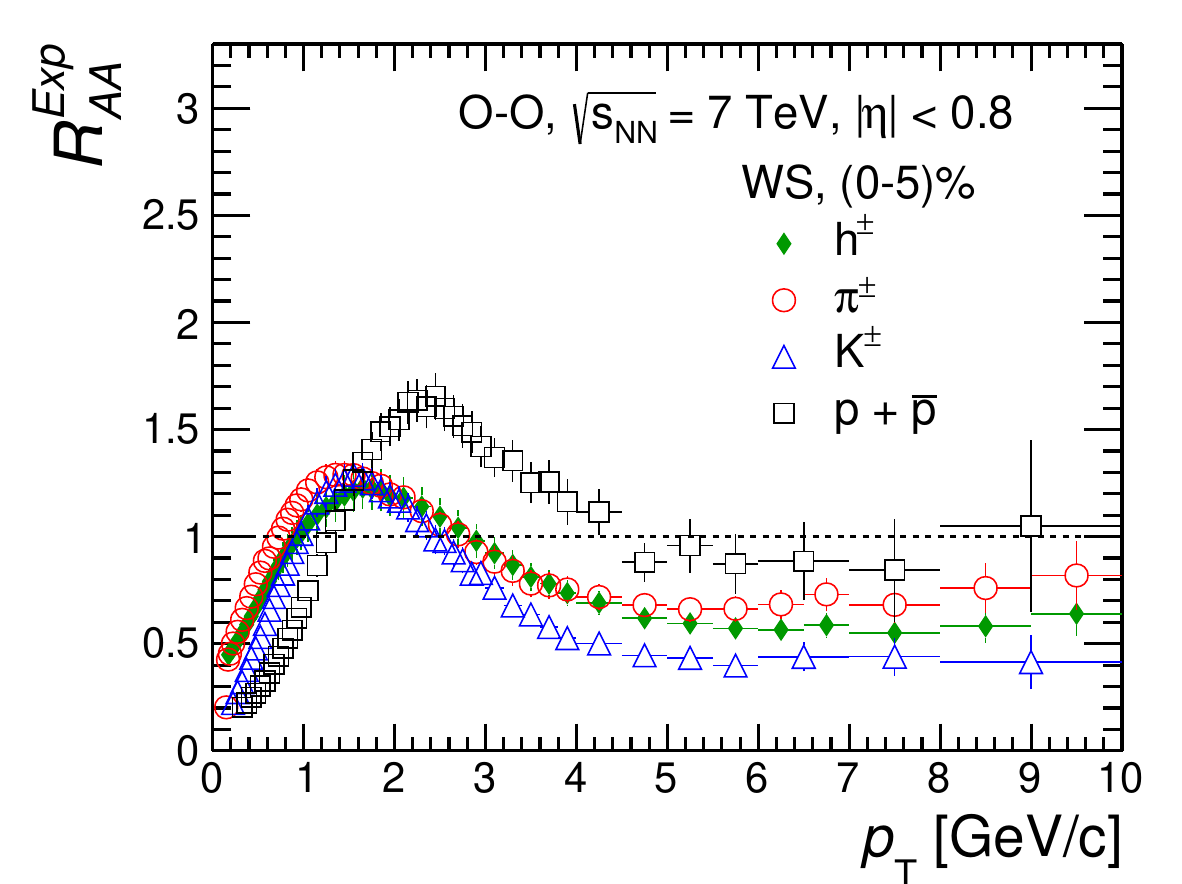}
\includegraphics[scale=0.295]{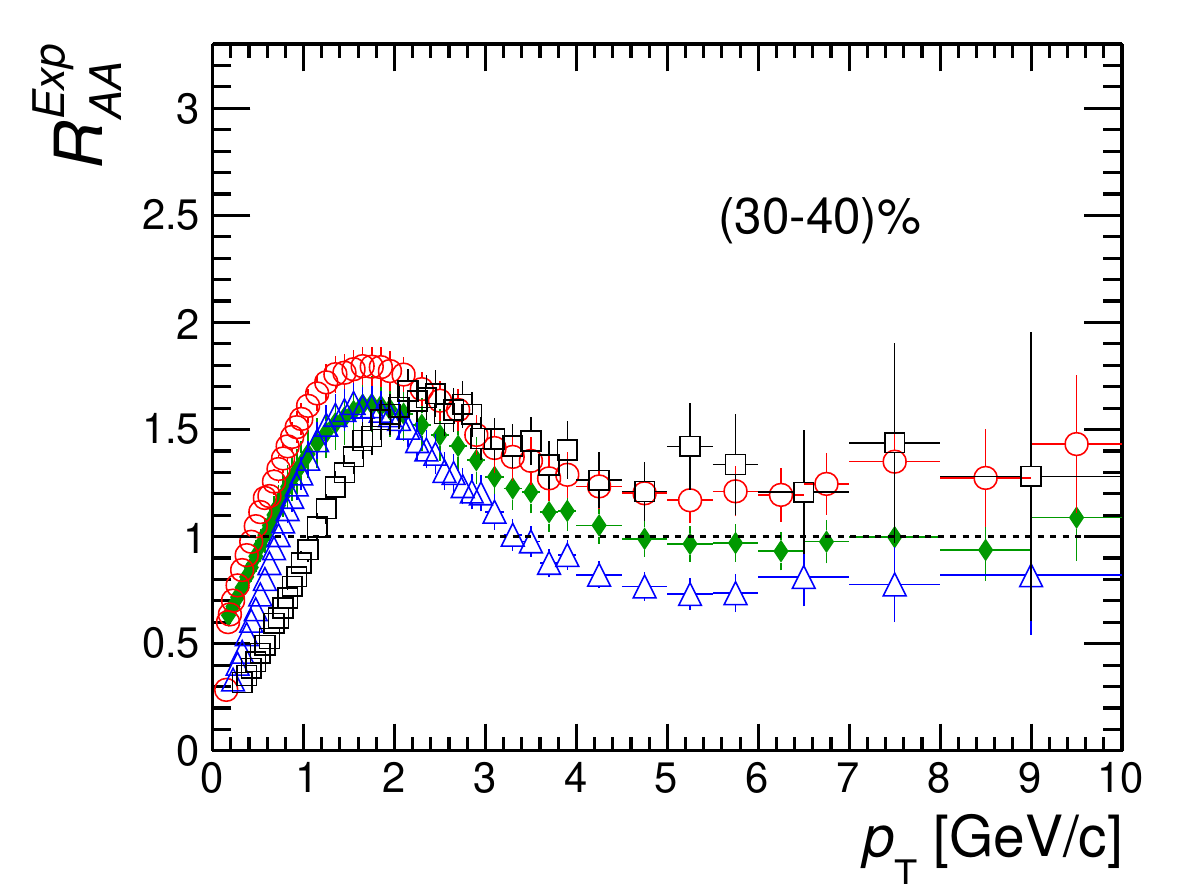}
\includegraphics[scale=0.295]{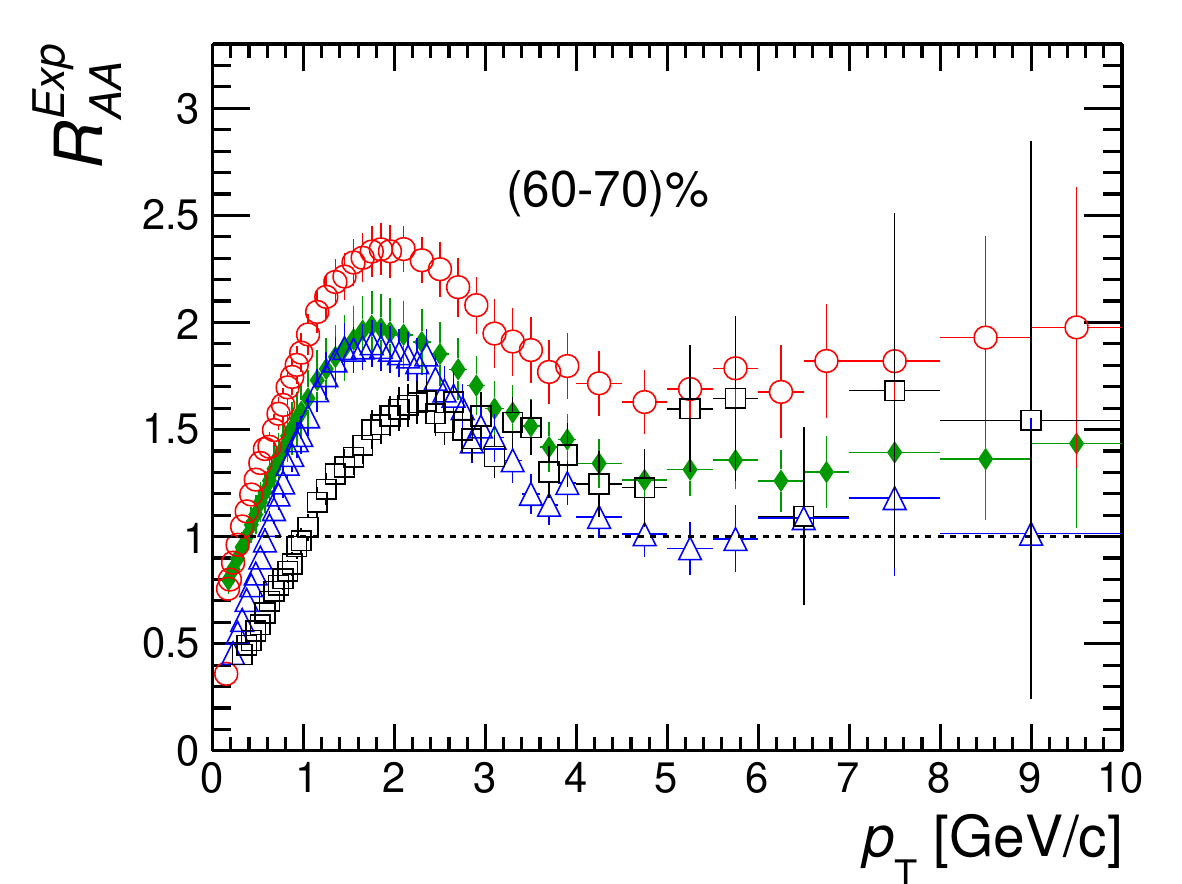}
\caption[]{(Color Online) Nuclear modification factor ($\textit R_{\textit {AA}}$) of all charged 
hadrons ($\rm{h}^{\pm}$) and identified particles ($\pi^{\pm}$, $\rm{K}^{\pm}$ and 
$\rm{p}+\bar\rm{p}$) in O-O collisions at $\sqrt{s_{\rm{NN}}}$ = 7 TeV for (0-5)\% 
[left], (30-40)\% [middle] and (60-70)\% [right], with Woods-Saxon density profile,  
taking $\textit{p}_{\rm{T}}$ spectra of $\textit{pp}$ from ALICE for estimating $\textit R_{\textit {AA}}$.}
\label{fig3}
\end{figure*}

\begin{figure*}[ht!]
\includegraphics[scale=0.295]{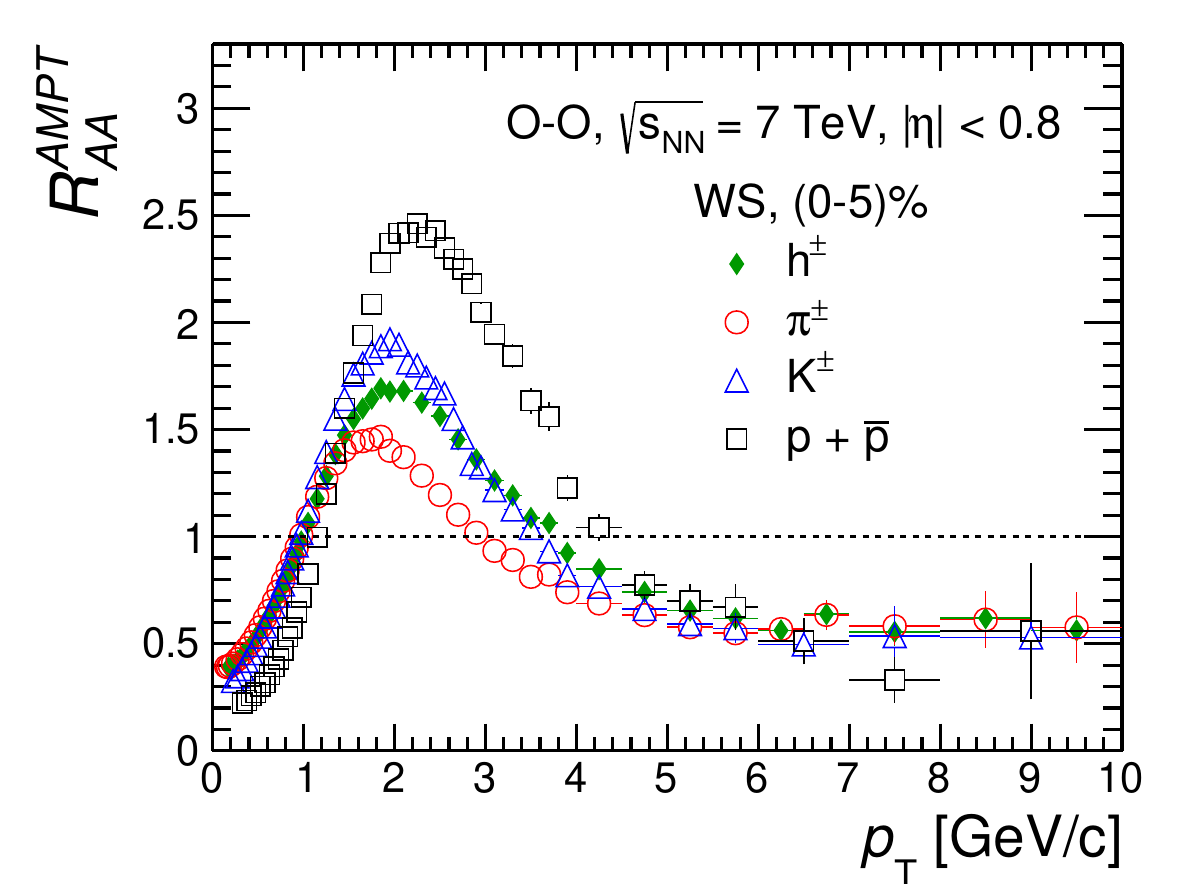}
\includegraphics[scale=0.295]{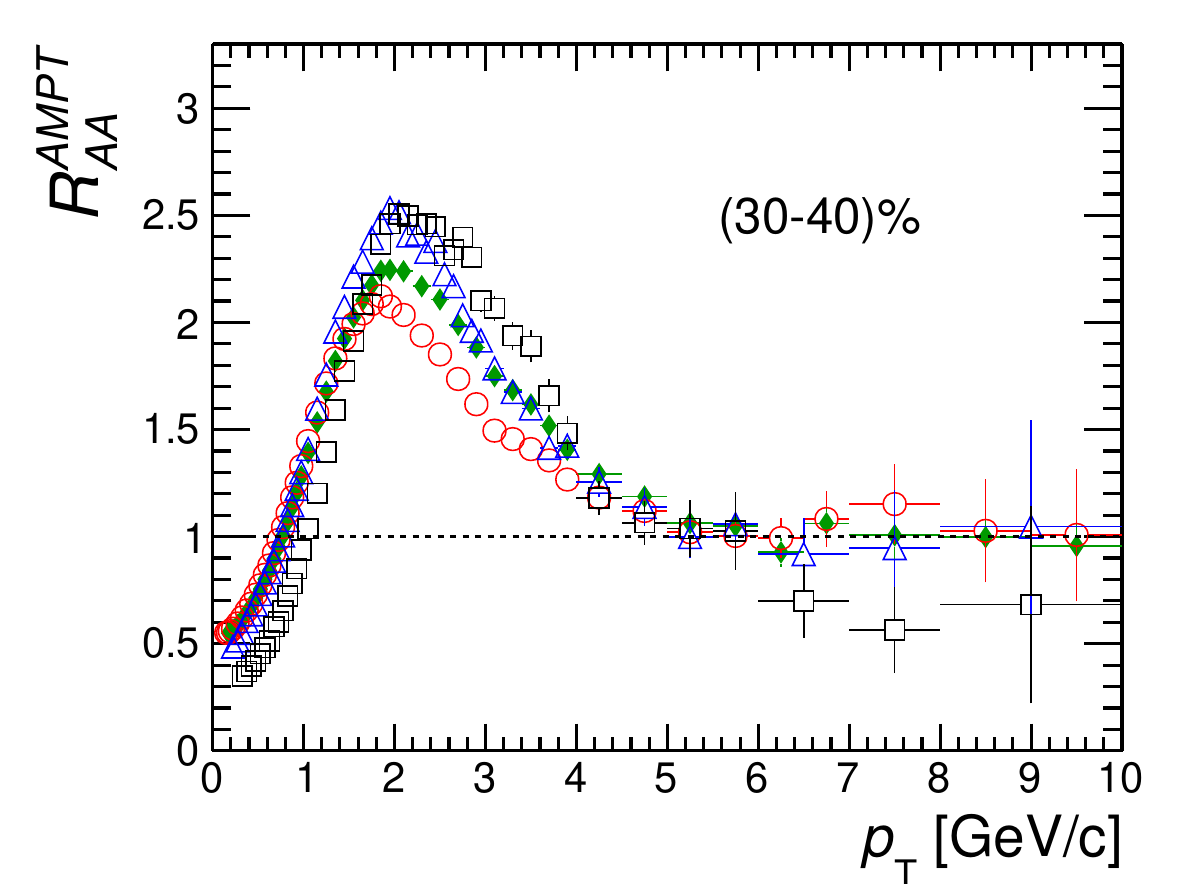}
\includegraphics[scale=0.295]{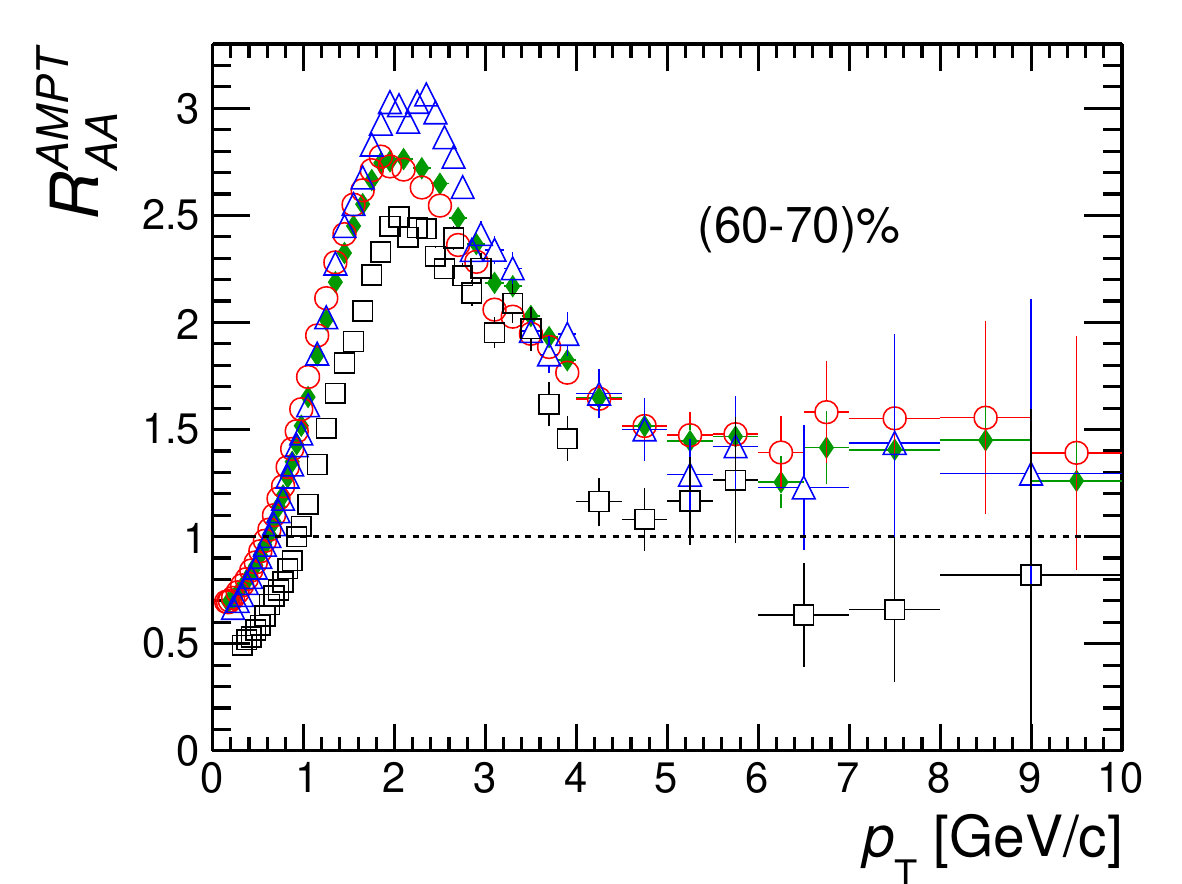}
\caption[]{(Color Online) Nuclear modification factor ($\textit R_{\textit {AA}}$) of all charged 
hadrons ($\rm{h}^{\pm}$) and identified particles ($\pi^{\pm}$, $\rm{K}^{\pm}$ and 
$\rm{p}+\bar\rm{p}$) in O-O collisions at $\sqrt{s_{\rm{NN}}}$ = 7 TeV for (0-5)\% 
[left], (30-40)\% [middle] and (60-70)\% [right], with Woods-Saxon density profile, 
taking $\textit{p}_{\rm{T}}$ spectra of $\textit{pp}$ from AMPT for estimating $\textit R_{\textit {AA}}$. }
\label{fig3ampt}
\end{figure*}

With the above comparison and validation of the simulation, we proceed to obtain 
$\textit R_{\textit {AA}}$ in ultra-relativistic O-O collisions. This system lies between Pb-Pb 
and $\textit{pp}$ collisions species with respect to the final state 
multiplicity~\cite{Huss:2020dwe}. Though, in the MC-Glauber model 
calculation, it is reported that the mean number of participants 
(${\rm<{N_{part}}>}$) in most  central (0-5)\% O-O collisions are comparable with 
the ${\rm<{N_{part}}>}$ in peripheral (60-70)\%  Pb-Pb 
collisions~\cite{Loizides:2017ack, Behera:2021zhi}. However, it has also been observed that the charged particle 
multiplicity obtained from  AMPT simulation of O-O collisions corresponding to the 
(0-5)\% centrality class 
approximately agrees with the (50-60)\% centrality of the Pb-Pb 
collisions~\cite{Behera:2021zhi}. Thus, this fact is expected to 
show up in the $\textit{p}_{\rm{T}}$ spectra as well. In Fig.~\ref{ptscomparision}, we have 
obtained charged particle 
$\textit{p}_{\rm{T}}$ spectra for the most central O-O collisions considering both 
$\alpha-$clustered and Woods-Saxon 
density profiles and compared with ALICE results corresponding to (50-60)\% and  (60-
70)\% centralities of Pb-Pb collisions at $\sqrt{s_{\rm{NN}}}$ = 5.02 TeV. It can be clearly seen from the 
bottom panel of 
Fig.~\ref{ptscomparision} that the results for O–O collisions at 
$\sqrt{s_{\rm{NN}}}$ = 7 TeV for (0-5)\% 
centrality agree well with the results for Pb-Pb collisions at $\sqrt{s_{\rm{NN}}}$ 
= 5.02 TeV for (50-60)\% 
centrality, within uncertainties, and for both the density profiles. Hence, in this 
work, we have chosen to 
make a comparative study between the most central O-O collisions (0-5)\% and the (50-60)\%  centrality class of Pb-Pb collisions.\\

\subsection{$\textit R_{\textit {AA}}$ vs. $\textit{p}_{\rm{T}}$  }
Prior to the estimation of $\textit R_{\textit {AA}}$ for the oxygen system, we further 
investigated the goodness of the AMPT model in the calculation of this observable in 
Pb-Pb collisions. This investigation using the model and comparison of ALICE results 
for all charged hadrons are depicted in Fig.~\ref{fig4}. These figures depict the 
$\textit{p}_{\rm{T}}$ dependence of $\textit R_{\textit {AA}}$ for both charged hadrons and identified 
particles in the (50-60)\% centrality class of Pb-Pb collisions at  
$\sqrt{s_{\rm{NN}}}$ = 5.02 TeV. Figure~\ref{fig4}, top (bottom) shows the 
estimation of $\textit R_{\textit {AA}}$ using the results of $\textit{pp}$ collisions from ALICE 
(AMPT) data. For both figures, the Pb-Pb yields are from the AMPT simulation. 
The comparison shows a substantial difference between simulation and 
experimental data towards $\textit{p}_{\rm{T}}  <  4 ~\rm GeV$. This discrepancy could be the 
consequence of the deviation observed in particle $\textit{p}_{\rm{T}}$ spectra towards 
low-$\textit{p}_{\rm{T}}$ range as seen in Fig.~\ref{pbpb50to60}. Further, at $\textit{p}_{\rm{T}} > 
3~$GeV, protons are observed to be less suppressed than other hadrons. The 
difference in species-specific suppression is compatible with a  mass ordering 
towards $\textit{p}_{\rm{T}}~<~ $1.5~GeV, implicating radial flow~\cite{ALICE:2019hno}. At~~$\textit{p}_{\rm{T}}~> 4$ GeV, 
a good agreement between experimental and AMPT models can be seen.\\

\begin{figure}[ht!]
\includegraphics[scale=0.45]{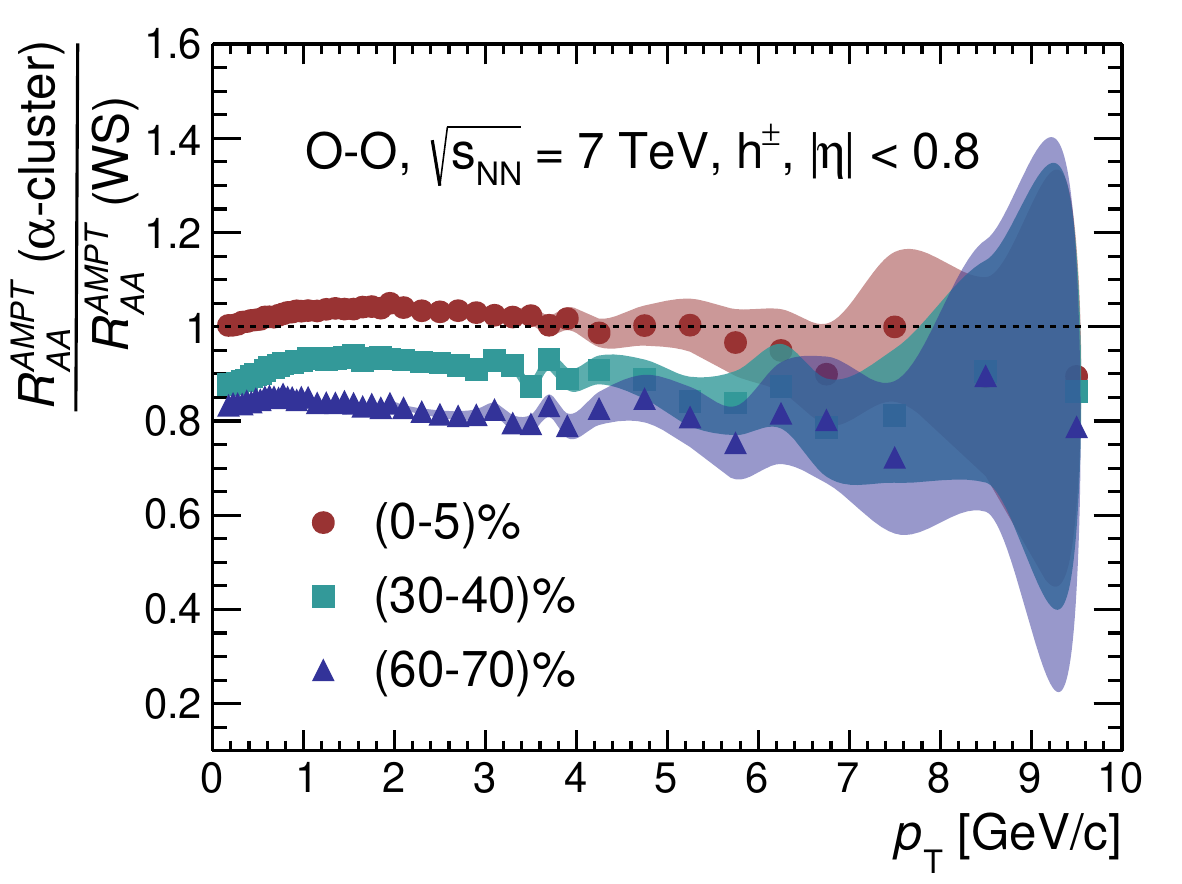}
\caption[]{(Color Online) Ratio between $\textit R_{\textit {AA}}$ of $\alpha-$clustered to Woods-Saxon density potential of all charged hadrons ($\rm{h}^{\pm}$) in O-O collisions at 
$\sqrt{s_{\rm{NN}}}$ = 7 TeV. The shaded region shows the statistical errors.}
\label{ratioalphaws}
\end{figure}

Now, we proceed to estimate $\textit R_{\textit {AA}}$ in O$-$O collisions at 
$\sqrt{s_{\rm{NN}}}$ = 7 TeV. However, to study the effect of the nuclear density 
profile on the production of charged particles, we considered both 
$\alpha-$clustered  (Figs.~\ref{fig2} and  \ref{fig2ampt}) and Woods-Saxon  
(Figs.~\ref{fig3} and  \ref{fig3ampt}) density profiles. Also, to reflect on the 
effect of scaling of $\textit R_{\textit {AA}}$ with $\textit{p}_{\rm{T}}$ spectra obtained from $\textit{pp}$ 
collisions, Fig.~\ref{fig2} and  \ref{fig2ampt} show results obtained considering 
scaling by ALICE experimental (AMPT simulated) data. The results are shown for the 
most central (0-5)\%, mid-central (30-40)\%, and peripheral (60-70)\% collisions. 
From these figures (\ref{fig2} - \ref{fig3ampt}), it is interesting to observe that 
a mass ordering between $\pi^{\pm}$, $\rm{K}^{\pm}$, and protons remains conserved 
toward $\textit{p}_{\rm{T}} < 2$ GeV, despite changes in density profiles and/or centralities. 
However, above $\textit{p}_{\rm{T}} > 2$ GeV, this pattern appears to break down for both 
density profiles. One can observe that $\textit R_{\textit {AA}}$ values estimated using $\textit{p}_{\rm{T}}$ spectra of $\textit{pp}$ collisions from AMPT (Fig.~\ref{fig2ampt}, \ref{fig3ampt}) 
shows a higher value than that obtained considering ALICE (Fig.~\ref{fig2}, 
\ref{fig3}). It is worth mentioning here that in line with the findings of 
ref.~\cite{ALICE:2017ban}, where mesons such as $\phi(1020)$ and K$^{*0}(892)$ 
exhibit smaller $\textit R_{\textit {AA}}$ values compared to protons at high-$\textit{p}_{\rm{T}}$, we 
observe a similar trend among the identified particles clearly for the most central 
collisions. This trend suggests a baryon-meson ordering. The consequences of density 
profiles for O$-$O collisions are further explored in Fig.~\ref{ratioalphaws}. Here, 
the ratio of $\textit R_{\textit {AA}}$ of all charged hadrons considering $\alpha-$clustered and 
Woods-Saxon density profiles are studied with transverse momentum at different 
centralities. It can be seen that at the most central collisions [(0-5)\%], the 
effects of $\alpha-$clustered and Woods-Saxon density profiles on charged hadrons 
yield are approximately the same. However, when transitioning from mid-central [(30-40)\%] to peripheral [(60-70)\%] collisions, considering the uncertainties, it 
becomes apparent that nuclei with oxygen and $\alpha-$clustered density profiles 
have a more significant impact on particle production compared to those with Woods-Saxon density profiles. This suggests that colliding nuclei with $\alpha-$clustered 
structure creates a compact and denser fireball, particularly in relatively non-central collisions in comparison with the Woods-Saxon density profile. This finding 
sheds light on the intricate relationship between the nuclei's internal structure, 
collision dynamics, and the subsequent particle production processes, contributing 
to a deeper understanding of the complex physics at play in O$-$O  collisions.\\

On further examination of Figs.~\ref{fig2}, \ref{fig2ampt} and 
Figs.~\ref{fig3},\ref{fig3ampt}, it is observed that in the intermediate $\textit{p}_{\rm{T}}$ 
range ($1.5 \rm~{GeV} < \textit{p}_{\rm{T}} < 3.0 $ GeV), $\textit R_{\textit {AA}}$ reaches its maximum 
before declining towards high-$\textit{p}_{\rm{T}}$. This decline could be attributed to a 
single energy loss mechanism~\cite{Liu:2022vbg, ALICE:2018vuu} that affects all 
particle species at high-$\textit{p}_{\rm{T}}$. In the low and mid-$\textit{p}_{\rm{T}}$ regions, 
radial boosts and $\textit{p}_{\rm{T}}$-broadening help to explain the observed 
trends~\cite{Cronin:1974zm}. Radial boosts cause particles with low-$\textit{p}_{\rm{T}}$ to 
move towards higher $\textit{p}_{\rm{T}}$ regions, resulting in reduced $\textit R_{\textit {AA}}$ in the 
low-$\textit{p}_{\rm{T}}$ region~\cite{Liu:2022vbg}. On the other hand, $\textit{p}_{\rm{T}}$-broadening 
caused by multiple parton interactions leads to a noticeable peak in the $\textit{p}_{\rm{T}}$ spectrum. Additionally, we observe that suppression is more pronounced in 
most central collisions compared to peripheral collisions. This could be attributed 
to the central collisions having relatively larger energy densities.\\

\begin{figure}[ht!]
\includegraphics[scale=0.45]{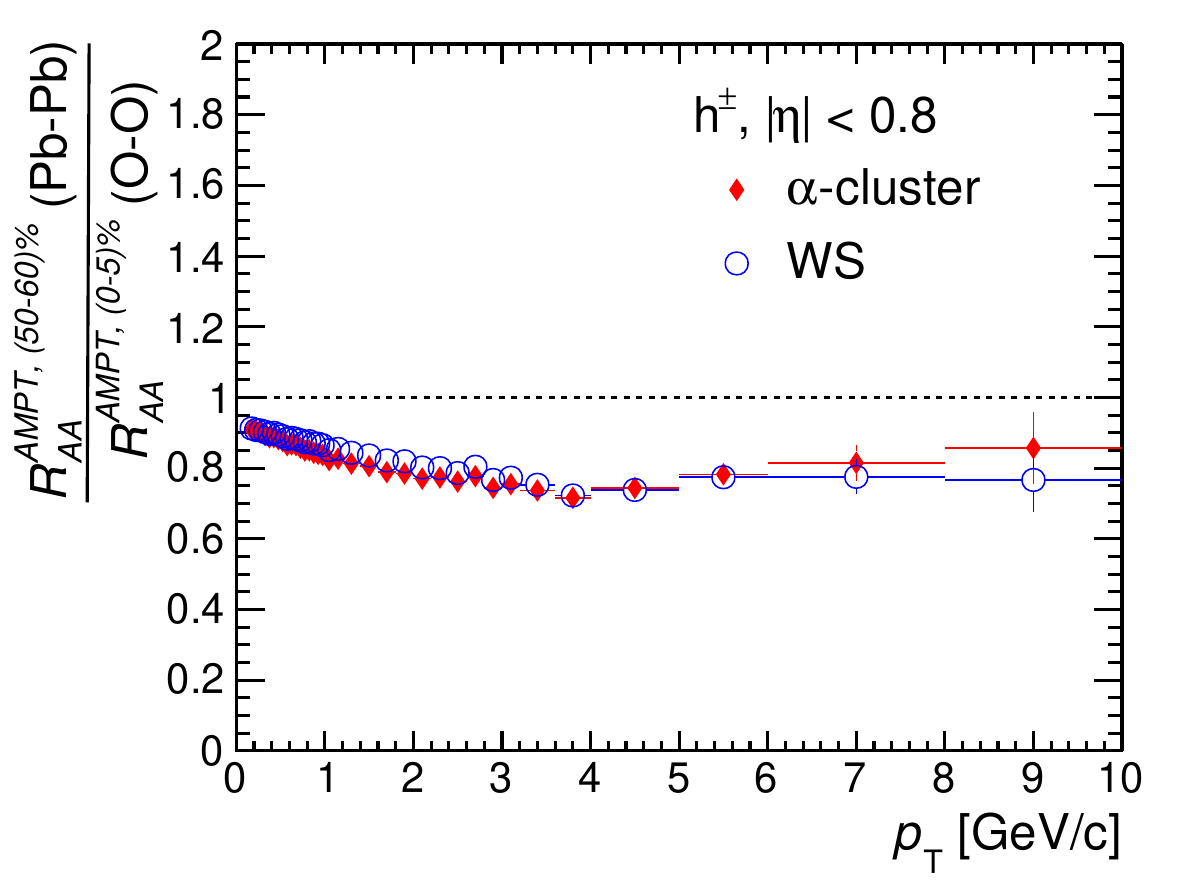}
\caption[]{(Color Online) Ratio between $\textit R_{\textit {AA}}$ in Pb-Pb collisions of (50-
60)\%  at $\sqrt{s_{\rm{NN}}}$ = 5.02 TeV to O-O collisions of (0-5)\%  at 
$\sqrt{s_{\rm{NN}}}$ = 7 TeV  for charged hadrons. Statistical errors are within the 
markers.}
\label{ratiopboo}
\end{figure}

It was discussed in Sec.~\ref{section3} that the (0-5)\% centrality class of O-O 
collisions and the (50-60)\% class of Pb-Pb collisions exhibit approximately similar 
multiplicities. To examine the consequences of this fact on the nuclear modification 
factor, we investigated the ratio of Pb-Pb (50-60)\% collisions to O-O (0-5)\% 
collisions for all charged hadrons, considering both $\alpha$-clustered and Woods-
Saxon density profiles for oxygen nuclei. This ratio is shown in 
Fig.~\ref{ratiopboo}, demonstrates that the $\textit R_{\textit {AA}}$ value in Pb-Pb collisions 
is smaller than in O-O collisions. Thus, the suppression effect is more pronounced 
in Pb-Pb collisions compared to O-O collisions within a similar multiplicity range. 
As observed in ref.~\cite{Sievert:2019zjr}, Pb-Pb collisions exhibit a 60\% larger 
radius than O$-$O collisions at a similar multiplicity. This observation implies 
that the partons traveling through the dense medium created in the Pb-Pb collisions 
(relatively larger nuclei) need to cover a longer path. As a result, this longer 
path length leads to increased energy loss when compared to O-O collisions. This 
observation is consistent with a study presented in Ref.~\cite{ALICE:2018hza}, where 
the $\textit R_{\textit {AA}}$ for Pb-Pb collisions and Xe-Xe collisions at similar multiplicity 
was examined.

\subsection{$\textit R_{\textit {AA}}$ variation with $\eta$}

\begin{figure}[ht]
\includegraphics[scale=0.45]{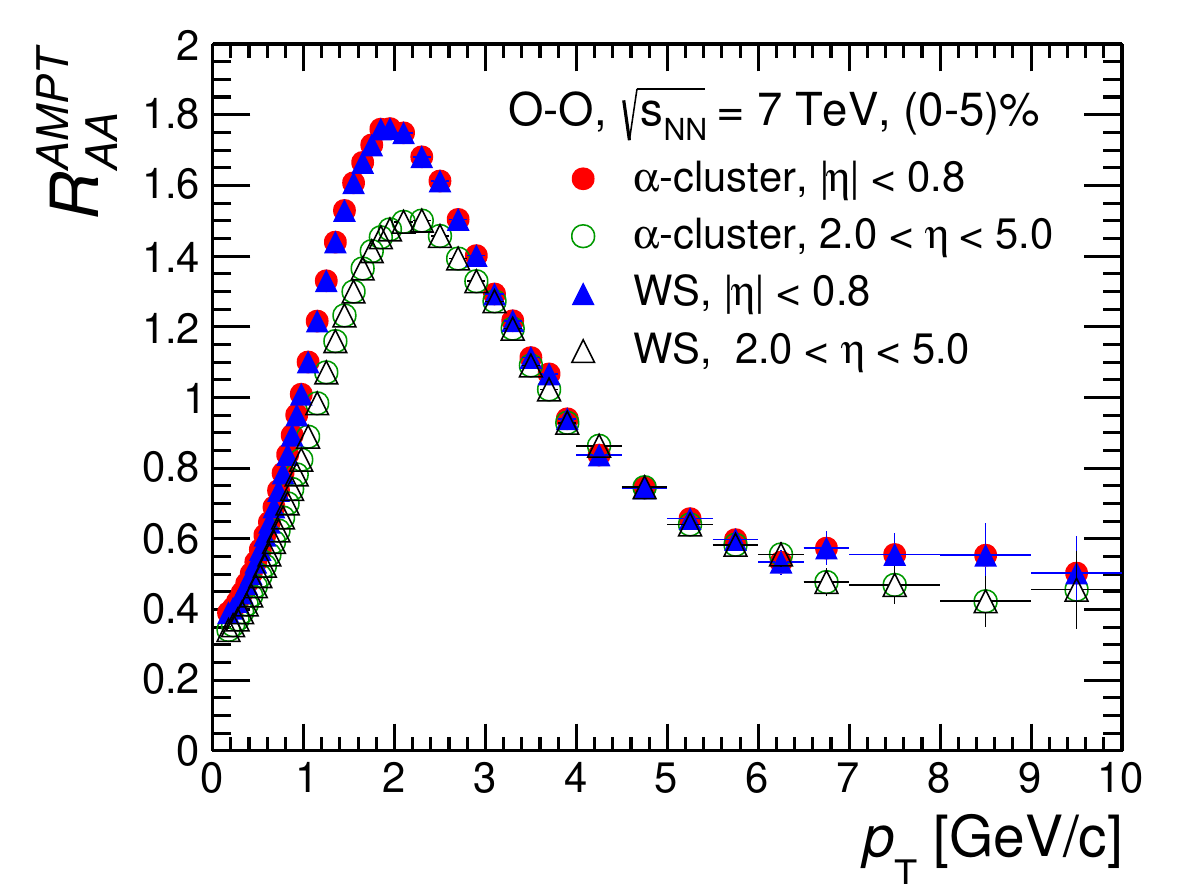}
\caption[]{(Color Online) Rapidity dependency of nuclear modification factor of 
charged hadrons in O-O collisions at $\sqrt{s_{\rm{NN}}}$ = 7 TeV for (0-5)\% 
centrality class.}
\label{fig5}
\end{figure}

As we know, the particle production at mid-rapidity is primarily because of the gluon-rich medium. While at forward rapidity, the particle production mechanism is dominated by the constituent quarks. Therefore, it is of interest to study the rapidity dependence on the relative yield of the particles produced in ultra-relativistic collisions. Here in Fig.~\ref{fig5}, we depicted  the $\textit R_{\textit {AA}}$ for all charged hadrons in (0-5)\% centrality class at midrapidity ($|\eta| < 0.8$) and forward-rapidity (2 $< \eta < 5 $) interval for both $\alpha$-cluster and Woods-Saxon density profiles in O-O collisions at $\sqrt{s_{\rm{NN}}}$ = 7 TeV. It is to be noted that the chosen rapidity regions have been taken from the Ref.~\cite{Behera:2021zhi}. The closed and open markers symbolize the nuclear modification factor obtained at midrapidity and forward rapidity, respectively. Figure~\ref{fig5} characterizes that the AMPT model predicts relatively less yield for all charged particles at forward rapidity compared with mid-rapidity up to $\textit{p}_{\rm{T}}\approx$ 3 GeV. However, at $\textit{p}_{\rm{T}}>3$ GeV, all the charged particles are suppressed with the same amount at mid and forward rapidity. It implies that particles moving at high speed are independent of the phase space distribution. It is consistent with the phenomena which suggest that the first moving particles escape the system before suffering any medium energy loss. We have also observed that the nuclear modification factor at a given rapidity is independent of Wood-Saxon and $\alpha$-cluster density profiles. \\

\section{Summary}
\label{section4}
This study investigates the nuclear modification factor ($\textit R_{\textit {AA}}$) for upcoming 
O-O collisions at $\sqrt{s_{\rm{NN}}}$ = 7 TeV at the LHC using a transport model. 
We consider the centrality of O-O collisions to be in a QCD environment similar to 
that of Pb-Pb collisions at $\sqrt{s_{\rm{NN}}}$ = 5.02 TeV, with respect to the 
final state charged-particle multiplicity. Additionally, we explore the dependence 
on density profiles, including Wood-Saxon and $\alpha$-clustered profiles. Finally, 
we study the rapidity dependence of the nuclear modification factor for O-O 
collisions at $\sqrt{s_{\rm{NN}}}$ = 7 TeV. The key findings, with regard to the O-O collisions are summarized below:\\

\begin{itemize}

\item  It is observed that the nuclear modification factor ($\textit R_{\textit {AA}}$) of 
charged hadrons and identified particles for O-O collisions have a smaller value for 
the most central (0-5)\% collisions in comparison to the peripheral collisions (60-70)\%, regardless of the density profiles used.\\

\item  It is observed that the mass-ordering among the identified particles remains 
consistent in the low-$\textit{p}_{\rm{T}}$ region ($<$ 2 GeV), irrespective of changes in 
density profiles and/or centralities. However, this trend ceases to exist for 
$\textit{p}_{\rm{T}} > 2$ GeV. In this context, a baryon-meson ordering becomes evident 
towards high-$\textit{p}_{\rm{T}}$ and is particularly prominent in most central collisions.

\item We have observed that the $\textit R_{\textit {AA}}$ values obtained using the $\textit{p}_{\rm{T}}$ spectra of $\textit{pp}$ collisions from the AMPT simulation exhibit higher values compared to those obtained using ALICE data. This shows that the production of hadrons is more in $\textit{pp}$ collisions at ALICE than in the AMPT simulation, highlighting the inadequacy of the AMPT model.

\item  From the study of $\textit R_{\textit {AA}}$ for charged hadrons and identified particles, 
we additionally observe that the effect of $\alpha$-clustered density profiles on 
particle production is more pronounced in mid-central and peripheral collisions 
compared to the most central collisions, as opposed to the Woods-Saxon density 
profile. This suggests that the nuclei with $\alpha$-clustered structure colliding 
may generate a more compact and denser fireball, particularly in relatively non-central collisions.

\item Furthermore, we observe that despite having the same final state multiplicity, 
the (50-60)\% centrality class of Pb-Pb collisions displays greater suppression than 
the (0-5)\% centrality class of O-O collisions. This underscores the significance of 
the system size to which the hadrons are exposed. Thus, here, Pb-Pb collisions 
create a larger nuclear environment compared to O-O collisions.

\item In the context of rapidity dependence,  we observed a relatively lower yield at forward rapidity compared to mid-rapidity for charged hadrons in O-O collisions at $\sqrt{s_{\rm{NN}}}$ = 7 TeV.
The nuclear modification factor in a given rapidity window is found to be independent of nuclear density profiles.

\end{itemize}

\section*{Acknowledgements}

DB acknowledges the support of a doctoral from CSIR, the Government of India. S.D. acknowledges the financial support from the postdoctoral fellowship of
CNRS at IJCLAB, Orsay, France. R.S. and C.R.S. acknowledge the financial 
support under the ALICE project (Project No. SR/MF/PS-02/2021-IITI (E-37123)). The 
usage of the ALICE Tier-3 computing facility at IIT Indore is gratefully 
acknowledged. The authors would like to acknowledge the help from Neelkamal Mallick 
related to the MC implementation of $\alpha$-clustering density profile in the 
Oxygen nucleus.

\end{document}